\documentclass[conference]{IEEEtran}

\usepackage{spconf}
\usepackage{graphbox}
\usepackage{array}
\usepackage{mdwmath}
\usepackage{mdwtab}
\usepackage{fixltx2e}
\usepackage{url}
\usepackage{color}
\usepackage[]{graphicx}
\usepackage{epsfig}
\usepackage{bbm}
\usepackage{enumerate}
\usepackage{amsfonts}
\usepackage{amsmath}
\usepackage{amssymb}
\usepackage{cite}
\usepackage{algorithm}
\usepackage{algpseudocode}
\usepackage{eqparbox}
\usepackage{comment}
\usepackage{caption}

\DeclareMathOperator*{\argmin}{argmin}

\IEEEoverridecommandlockouts

\begin{document}
\title{DEEP NEURAL NETWORKS FOR NON-LINEAR MODEL-BASED ULTRASOUND RECONSTRUCTION}



\name{H. Almansouri$^{\star}$   S.V. Venkatakrishnan$^{\dagger}$  G.T. Buzzard$^{+}$  C.A. Bouman$^{\star}$  H. Santos-Villalobos$^{\dagger}$}

\address{$^{\star}$ School of Electrical and Computer Engineering, Purdue University, West Lafayette, IN, 47907 \\
    $^{\dagger}$Imaging, Signals and Machine Learning Group, Oak Ridge National Laboratory, Oak Ridge, TN, 37831\\
    $^{+}$Department of Mathematics, Purdue University, West Lafayette, IN, 47907}

\maketitle

\begin{abstract}
	
	Ultrasound reflection tomography is widely used to image large complex specimens that are only accessible from a single side, such as well systems and nuclear power plant containment walls.
	Typical methods for inverting the measurement rely 
	on delay-and-sum algorithms that rapidly produce reconstructions but with significant artifacts. 
	Recently, model-based reconstruction approaches using a linear forward model have been shown to significantly improve image quality compared to the conventional approach. 
	However, even these techniques result in artifacts for complex objects because of the inherent non-linearity of the ultrasound forward model. 
	
	In this paper, we propose a non-iterative model-based reconstruction method for inverting measurements that are based on non-linear forward models for ultrasound imaging.  
	Our approach involves obtaining an approximate estimate of the reconstruction using a simple linear back-projection and training a deep neural network to refine this to the actual reconstruction. 
	We apply our method to simulated and experimental ultrasound data to
	demonstrate dramatic improvements in image quality compared to the delay-and-sum approach and the linear model-based reconstruction approach. 
	
	\vspace{-0.2in}
\end{abstract}
{\let\thefootnote\relax\footnote{{This manuscript has been authored by UT-Battelle, LLC under Contract No. DE-AC05-00OR22725 with the U.S. Department of Energy.  The United States Government retains and the publisher, by accepting the article for publication, acknowledges that the United States Government retains a non-exclusive, paid-up, irrevocable, world-wide license to publish or reproduce the published form of this manuscript, or allow others to do so, for United States Government purposes.  The Department of Energy will provide public access to these results of federally sponsored research in accordance with the DOE Public Access Plan (http://energy.gov/downloads/doe-public-access-plan).}}}

\section{Introduction}

\begin{figure}[!htbp]
	\begin{center}
		\includegraphics[scale=0.2]{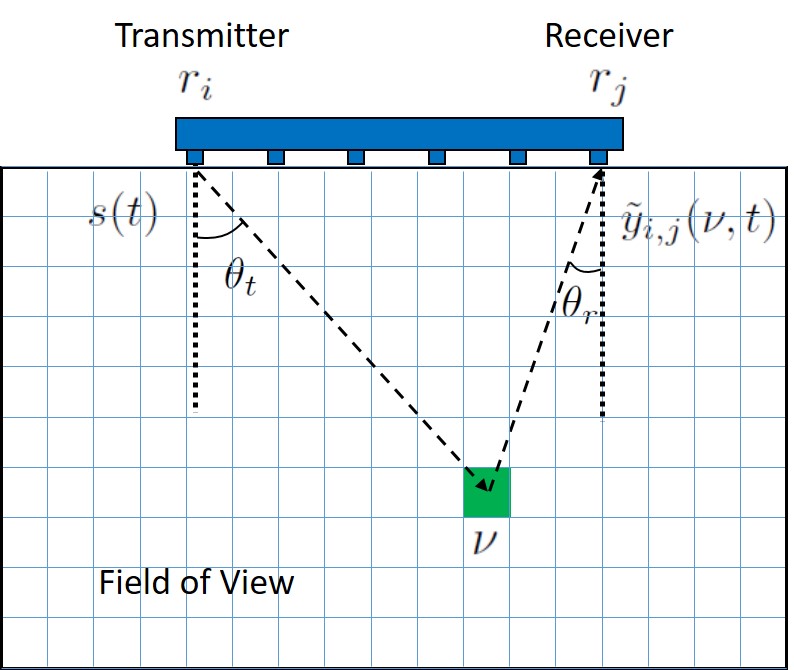}
	\end{center}
	\caption{Illustration of a typical ultrasound system for non-destructive evaluation. The transducers are used to make pulse-echo measurements which are processed to reconstruct the cross-section.}
	\label{fig:us_system}
	\vspace{-0.1in}
\end{figure}

One-sided ultrasound reflection tomography is vital for non-destructive evaluation (NDE) of large heterogeneous specimens, such as the casing of injection wells and thick concrete walls \cite{hoegh2015extended}. 
A typical system uses an array of transducers to transmit a signal from one sensor and receive at the others (see Fig.~\ref{fig:us_system}). 
The collection of received timeseries signals is then processed to reconstruct a cross-section of the object being imaged. 
Due to the need for rapid reconstructions, full waveform inversion approaches \cite{bernard2017ultrasonic} are not practical for ultrasound NDE and hence analytic algorithms based on a delay-and-sum approach, such as the synthetic aperture focusing technique (SAFT), are routinely used for reconstructions of ultrasound reflection mode data\cite{shao2011,Engle2014,dobie2013}.
Recently, we have developed a model-based iterative reconstruction \cite{almansouri2018anisotropic} approach using a simplified linear model and demonstrated significant improvements in reconstruction performance compared to SAFT while still being able to produce a reconstruction in near real-time.  
However, this linear-MBIR (L-MBIR) method still results in artifacts, such as reverberation and shadowing, due to the inherent non-linearity of the ultrasound system.
In summary, existing approaches for ultrasound reflection imaging for NDE may result in reconstructions with significant artifacts. 

There have been several recent efforts to use deep convolutional neural networks (CNN) to address inverse problems in imaging \cite{mccann2017convolutional}. 
One class of algorithms applies a two-step, non-iterative approach composed of a simple inversion followed by a CNN to obtain a reconstruction 
for inverse problems such as 
tomography \cite{han2017deep,jin2017deep}, MRI \cite{han2018deep,wang2016accelerating}, photo-acoustic tomography \cite{antholzer2017deep,hauptmann2018model}, compressed sensing \cite{mousavi2017learning}, and  non-linear optical imaging based on multiple scattering\cite{sun2018efficient}.  
Alternatively, researchers have adapted variable splitting strategies 
such as the Plug-and-Play approach \cite{venkatakrishnan2013plug,sreehari2016plug}
to iteratively solve two \textit{learned} sub-problems corresponding to a forward-model inversion and a denoising step in order to determine a fixed point  \cite{zhang2017learning,gupta2018cnn,rick2017one,hauptmann2018model,adler2018learned,meinhardt17learning}. 
In summary, deep-learning based techniques have demonstrated promising results for a variety of inverse problems in imaging. 

In this paper, we propose a learning based approach for ultrasound reflection mode imaging using a non-iterative two-stage strategy.
Since the underlying forward model is non-linear, we first obtain a preliminary reconstruction based on the adjoint of a simple \textit{linear} model for the ultrasound system. 
We then train a multi-scale deep convolutional neural network to map this initial reconstruction to the true reconstruction.
Importantly, the CNN can account for the nonlinear and space varying effects in the ultrasound forward model, as well as account for attributes of the prior model that can be used to suppress image artifacts and noise. 
Once the network is trained, the algorithm can be applied in real time, because both steps can be performed rapidly using GPUs.
We demonstrate that the proposed approach dramatically improves image quality for ultrasound imaging compared to SAFT and L-MBIR by removing artifacts caused by the non-linearity of the system, such as reverberation and shadowing artifacts. The improvements are more evident for image targets buried deep inside the object being inspected. 
Also, the proposed CNN-based approach is able to reconstruct the specimen's acoustic speeds from the back-projected timeseries signals, yielding a quantitative reconstruction compared to existing approaches that are qualitative.

The organization of the rest of this paper is as follows. 
In section \ref{sec:us_forward} we introduce the ultrasound forward 
model and the conventional linear model used for inversion. 
In section \ref{sec:deep_learn} we present details of the proposed 
inversion algorithm based on a deep neural network. 
Finally in section \ref{sec:results} we present our results based on simulated phantom data.

\section{Ultrasound Forward Model and Linear Model-Based Inversion }
\label{sec:us_forward}

\begin{figure}[!htbp]
	\begin{center}
		\includegraphics[scale=0.25]{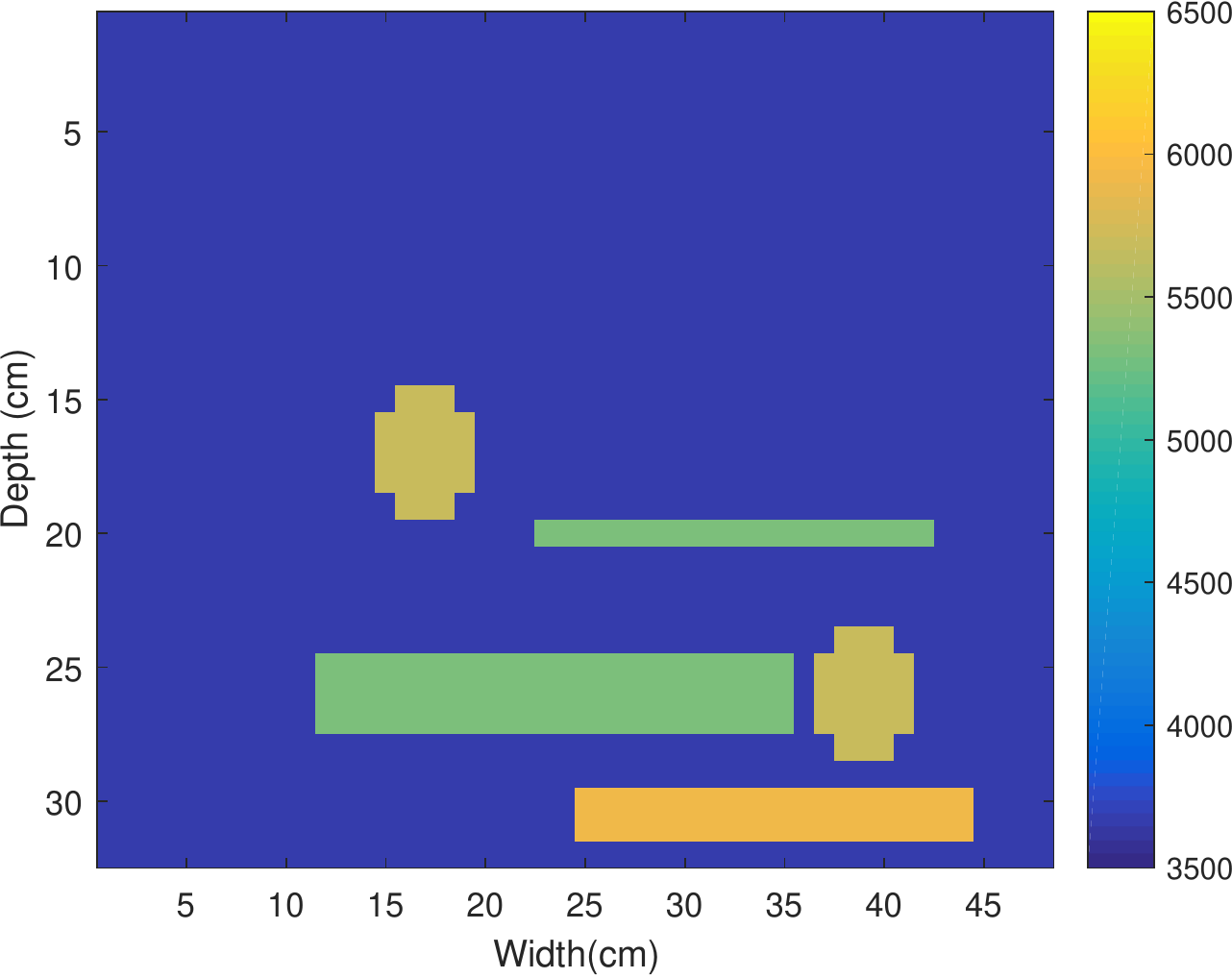}
		\includegraphics[scale=0.25]{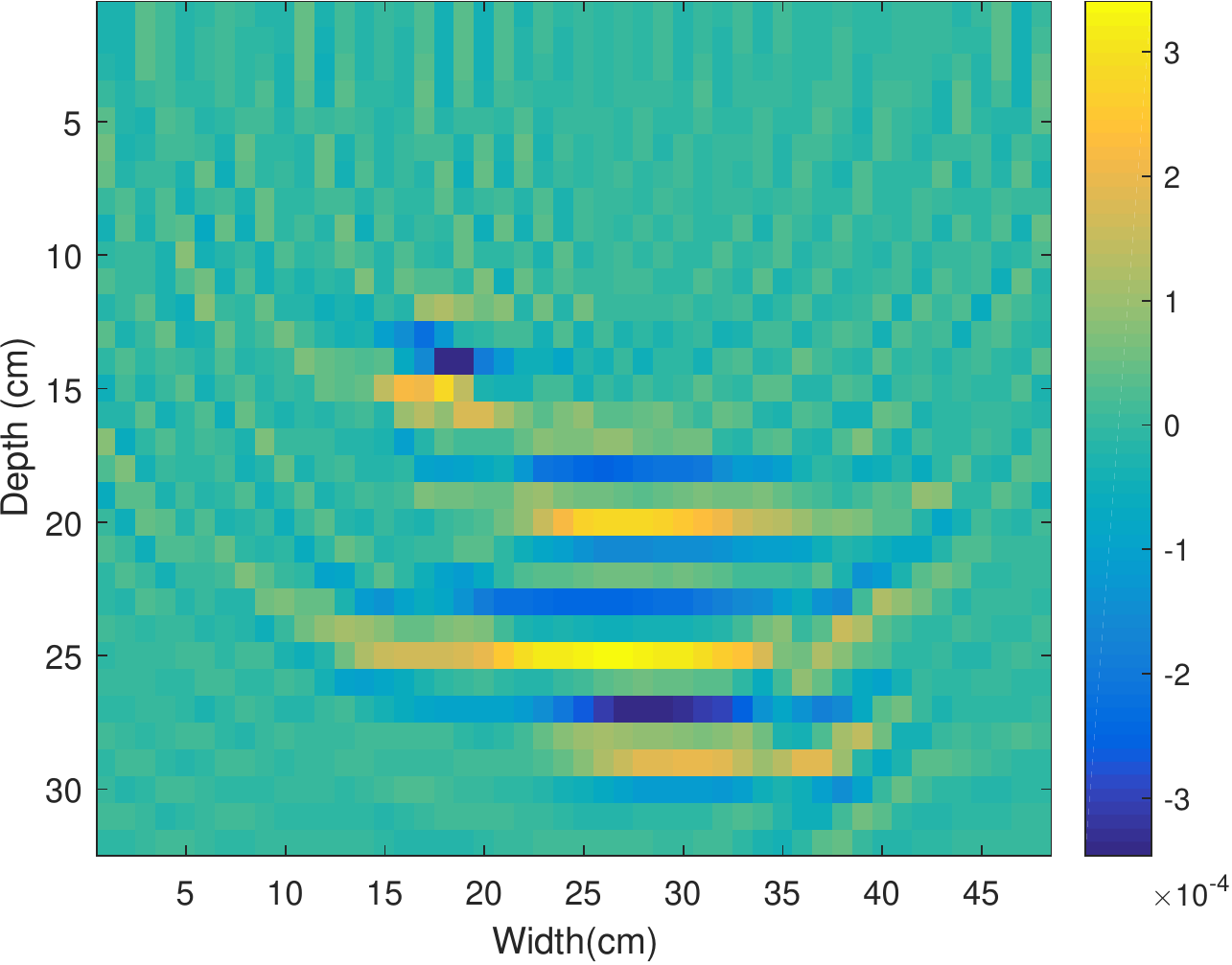}
	\end{center}
	\caption{Illustration of a back-projection of a one-sided ultrasonic NDE measurements using the system matrix $A$ in Eq. \ref{eq:us_lin}. The left image is the ground truth (speed-of-sound in units of m/s) and the right image is the back-projection of the simulated measurements obtained from the ground truth using an array of $10$ transducers and a non-linear wave propagation model. The back-projection suffers from artifacts and does not faithfully reconstruct the object.}
	\label{fig:BackP}
\end{figure}

\begin{figure*}[!htbp]
	\begin{center}
		\includegraphics[scale=0.35,clip, trim=0cm 5.4cm 0cm 3cm]{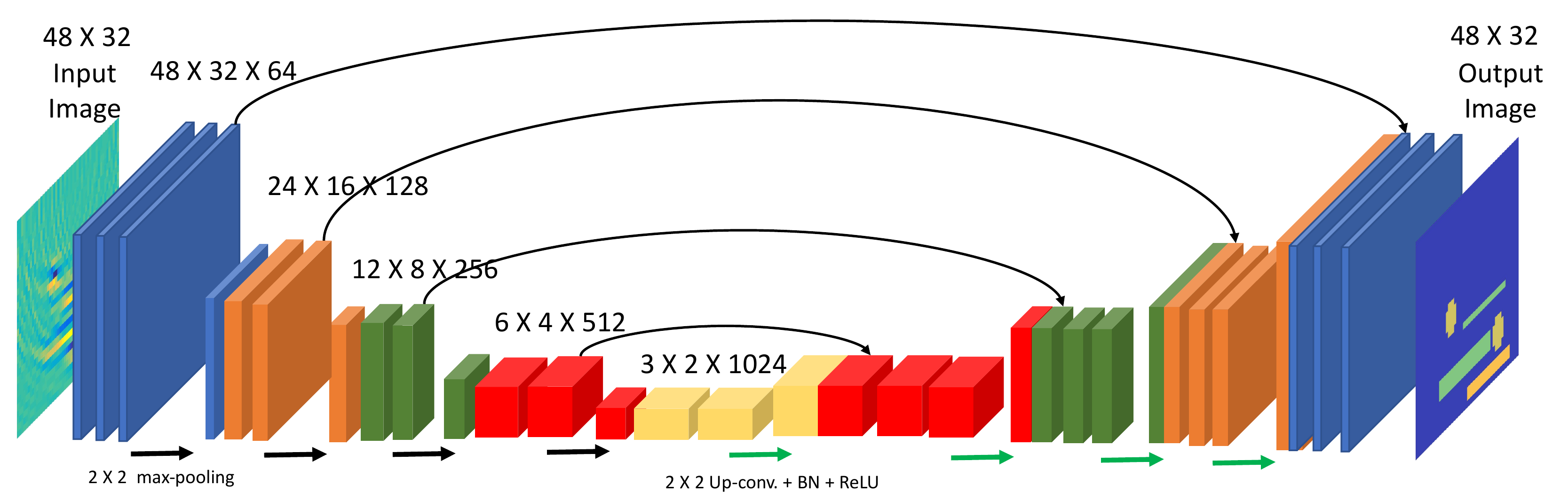}
	\end{center}
	\caption{Modified U-net architecture used for the reconstructions. The input is an image obtained by applying the adjoint of a linear operator to the measurements. Within each stage, we apply a $3 \times 3$ convolution followed by a batch normalization and a rectified linear unit. The size of the feature maps at each stage is noted in the image.}
	\label{fig:unet_arch}
\end{figure*}

The goal of ultrasound reflection mode imaging is to determine the properties of a cross-section being imaged using a transducer array (See Fig.~\ref{fig:us_system}).
In particular, the ultrasound wave propagation in a medium can be described by a set of coupled partial differential equations \cite{kwave2010},
\begin{eqnarray}
\frac{\partial u}{\partial t} &=& - \frac{1}{\rho_{0}} \nabla p,\nonumber\\ 
\frac{\partial \rho}{\partial t} &=&  -\rho_0 \nabla u - u \nabla \rho_{0},\nonumber\\ 
p &=&c_{0}^2 (\rho + d \cdot \nabla \rho_0 - L \rho),   \label{eq:us_nonlin}
\end{eqnarray}
where $u$ is the acoustic particle velocity, $\rho$ is the acoustic density, $d$ is the acoustic particle displacement, $L$ is an operator defined by
$$
L = - 2 \alpha_0 c_0^{y-1} \frac{\partial}{\partial t}\left( - \nabla^2 \right)^{\frac{y}{2}-1} + 2 \alpha_0 c_0^y \tan\left(\frac{\pi y}{2}\right) ( - \nabla^2 )^{\frac{y+1}{2}-1},
$$
and $0 < y < 3, y \neq 1$ is a parameter that controls the behavior of the absorption and dispersion. 
For the forward (simulation) model, the inputs to  this system of equations are the 2D fields corresponding to $c_0$, the acoustic velocities; $\rho_0$, the ambient densities; and $\alpha_0$, the attenuation.
The output is the pressure $p$ measured at the locations $r_j$ of the sensors as a function of time, $t$.
These measurements are then concatenated to form the measurement vector $y$.
Abstractly, we can represent this forward model relationship as
$$
y = f(c_0, \rho_0, \alpha_0 ) \ .
$$
Using these equations we can solve for the pressure at the sensor locations for a given input signal in order to simulate the received signal. 
However, the inversion of the underlying quantities from the received signals based on this model is challenging because of the complicated and non-linear nature of the forward model.

To address these challenges, we developed a simplified  linear model for the measurements \cite{almansouri2018anisotropic}, given by 
\begin{eqnarray}
\tilde{y}_{i,j} (t) =  \int_{\mathbb{R}^3} \tilde{A}_{i,j}(\tau_{i,j}(\nu),t) \tilde{x}(\nu) d\nu + \tilde{d}_{i,j}(t), \label{das}	
\end{eqnarray}
where $\tilde{y}_{i,j}$ is the measurement at the transmit-receive pair $(i,j)$, $\nu$ is a point in the field of view, $\tilde{A}_{i,j}$ is a response function that accounts for the time-shift and attenuation of the transmitted pulse, $\tilde{x}$ is the reflection coefficient, $\tau_{i,j}$ is the time delay of the transmitted signal for point $\nu$ and the measurement pair $(i,j)$, $\tilde{d}_{i,j}$ is the direct arrival signal.
Using this model, we designed a fast model-based reconstruction approach \cite{almansouri2018anisotropic} (L-MBIR) which works by minimizing the cost-function
\begin{eqnarray}
\hat{v} \leftarrow \argmin_{v} \left\{\frac{1}{2}||y-Av||_2^2 + R(v) \right\},
\label{eq:us_lin}
\end{eqnarray}
where $A$ is a projection matrix which discretizes $\tilde{A}$, $v$ is a vector of reflection coefficients and $R$ is a Markov random field based regularizer \cite{thibault2007three}. 
While this model is simple and significantly improves the reconstructions compared to conventional delay-and-sum approaches like SAFT, the method can result in artifacts in the reconstructed images due to the assumption of linearity. 
Furthermore, the reflection coefficient may not have a clear quantitative interpretation compared to quantities such as the speed, density or attenuation in the medium. 

\section{Deep Neural Network for Non-Linear Ultrasound Inversion }
\label{sec:deep_learn}
Since computing the exact solutions to \eqref{eq:us_nonlin} is expensive, we propose a two stage approach to the inversion. 
In the first step we leverage our previously introduced linear model and use the $A$ matrix in \eqref{eq:us_lin} to estimate an initial reconstruction $\tilde{v}=A^{T}y$. 
While this method highlights some of the essential features, such a reconstruction is not quantitative and has severe artifacts due to the non-linearity in the system (see Fig.~\ref{fig:BackP}).
In order to compensate for these artifacts, we use such an image as input to a deep-neural network that has been trained to map such an input to the actual image of the desired material properties such as the speed of sound in the medium. 
In particular, we use the U-net \cite{ronneberger2015u} with skip-connections to learn a mapping of this initial image to the actual reconstruction (see Fig.~\ref{fig:unet_arch}). 
This architecture is desirable because it has the entire input image in its receptive field and can hence learn features that are globally correlated. 
Furthermore the presence of skip-connections ensures that the architecture combines the features from different scales effectively.
We will refer to the proposed technique as direct deep learning (DDL) for the rest of the paper.

\section{Results}
\label{sec:results}

\begin{figure}[!t]
	\begin{center}
		\includegraphics[scale=0.27]{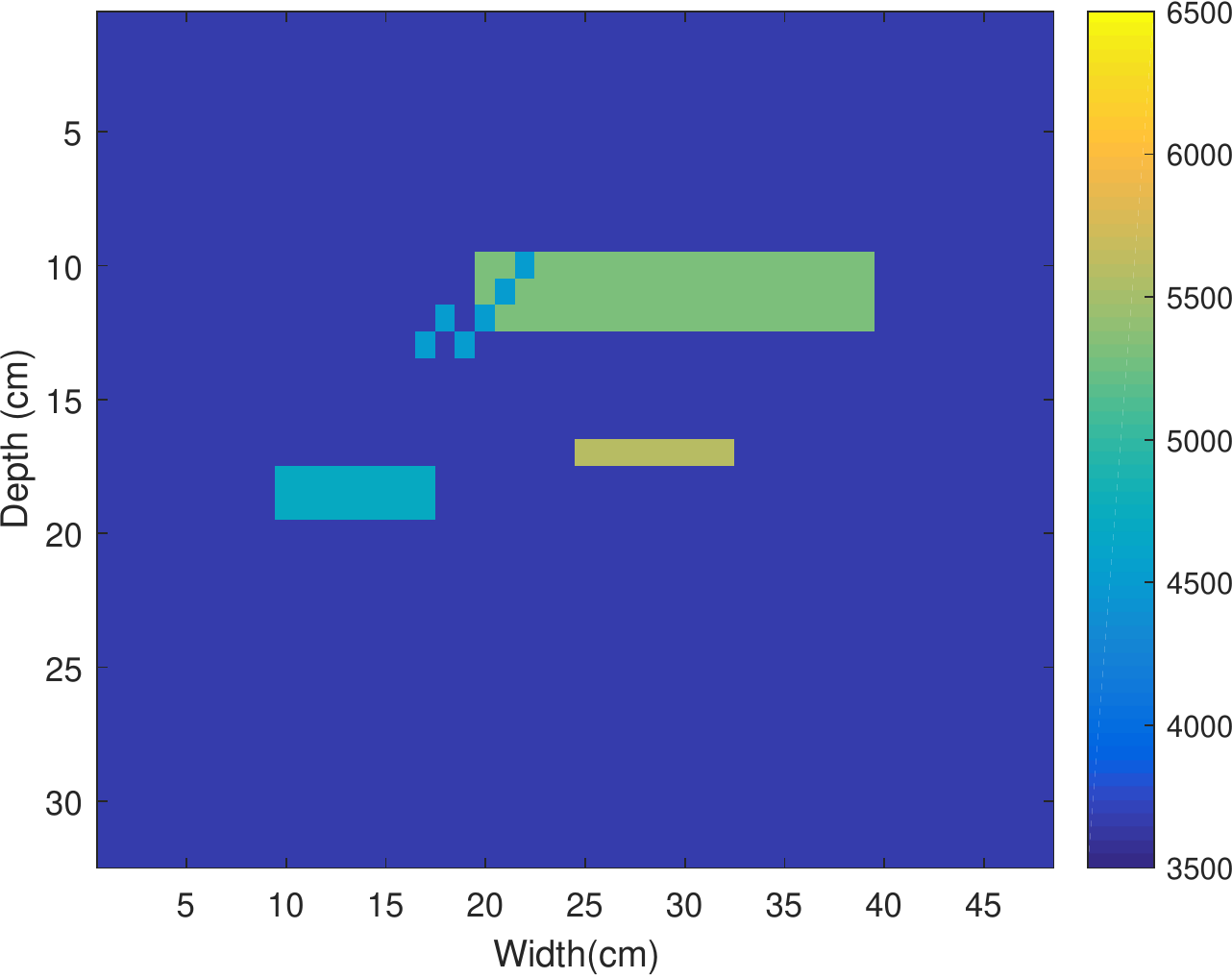} 
		\includegraphics[scale=0.27]{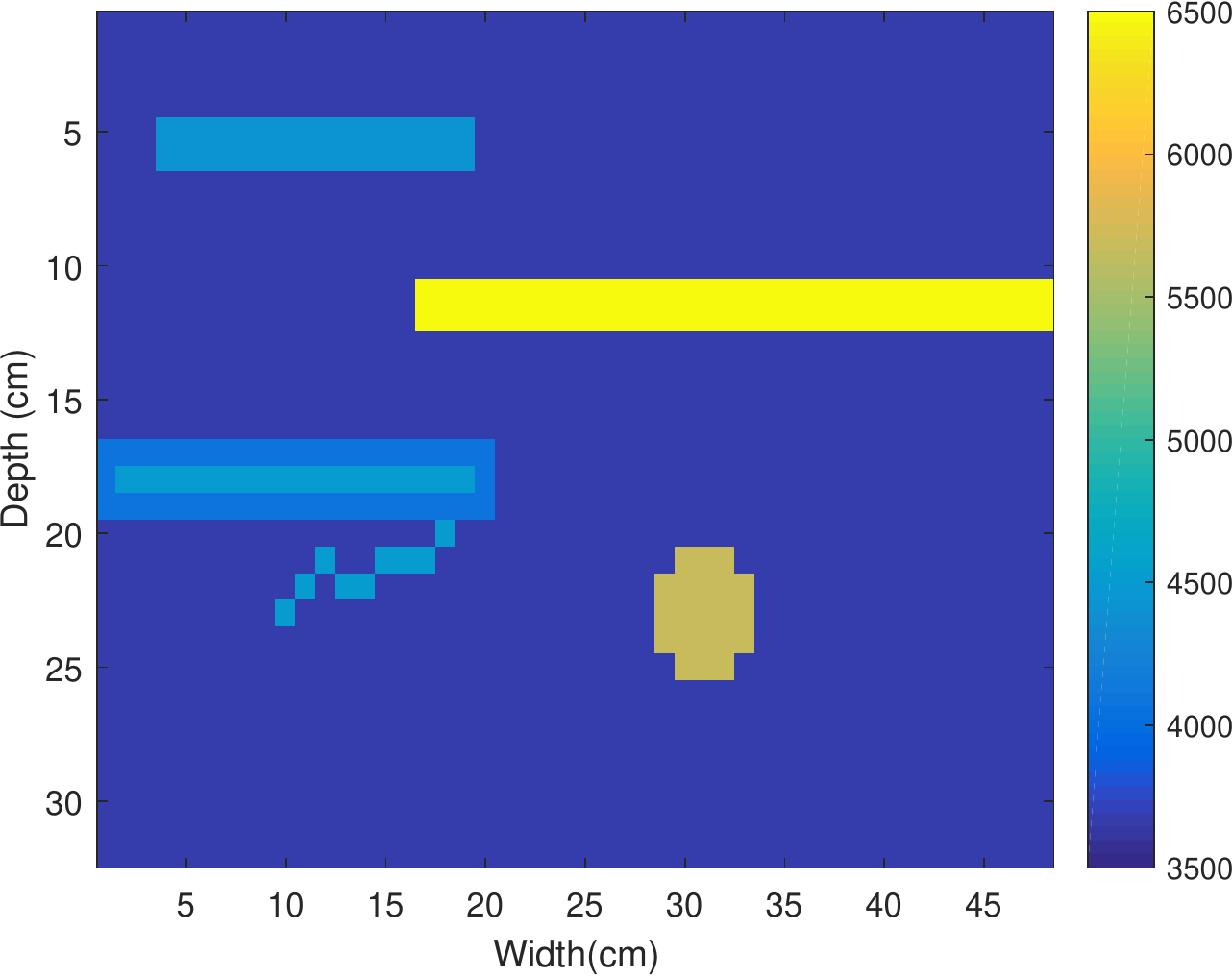} \\
		\includegraphics[scale=0.27]{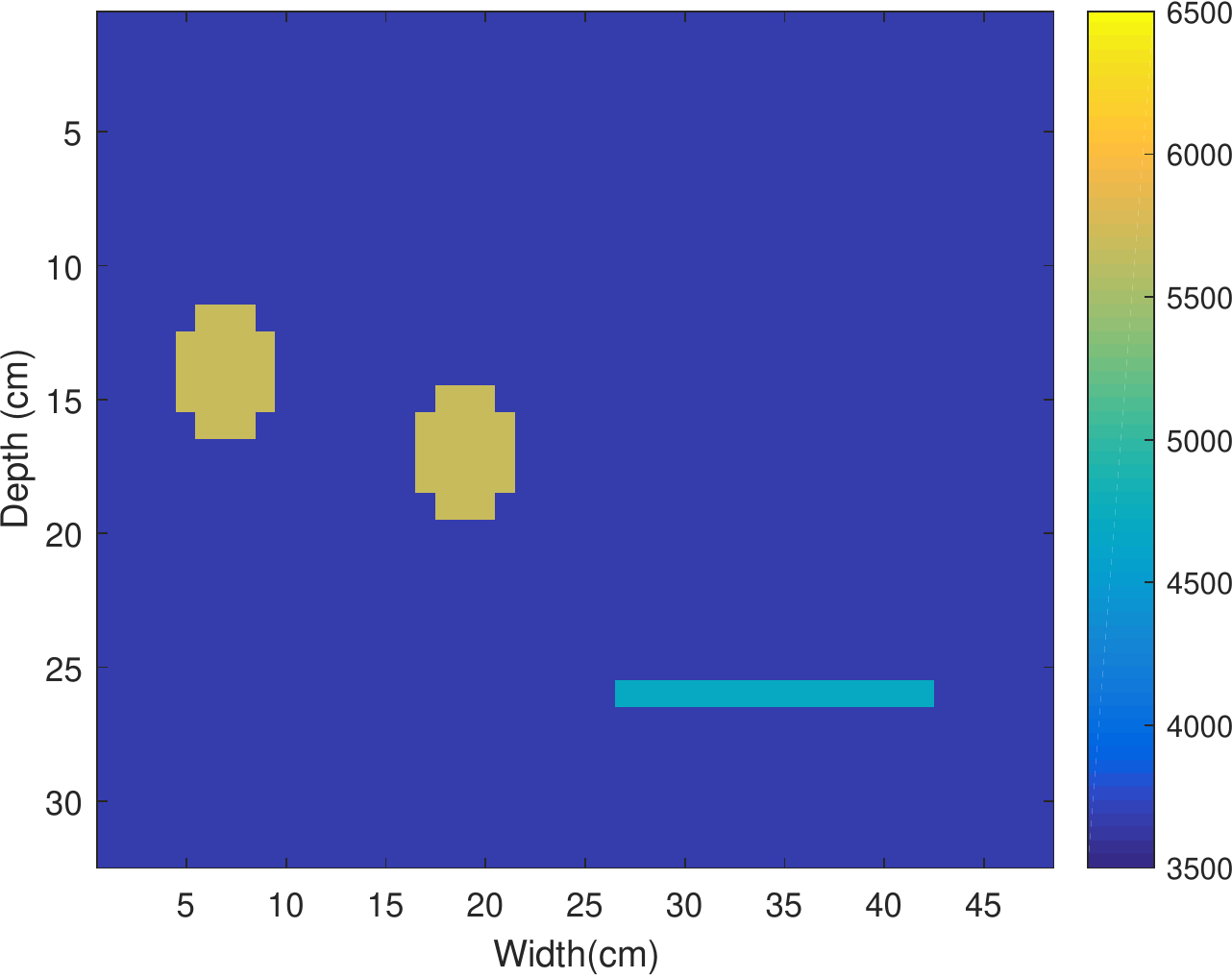} 
		\includegraphics[scale=0.27]{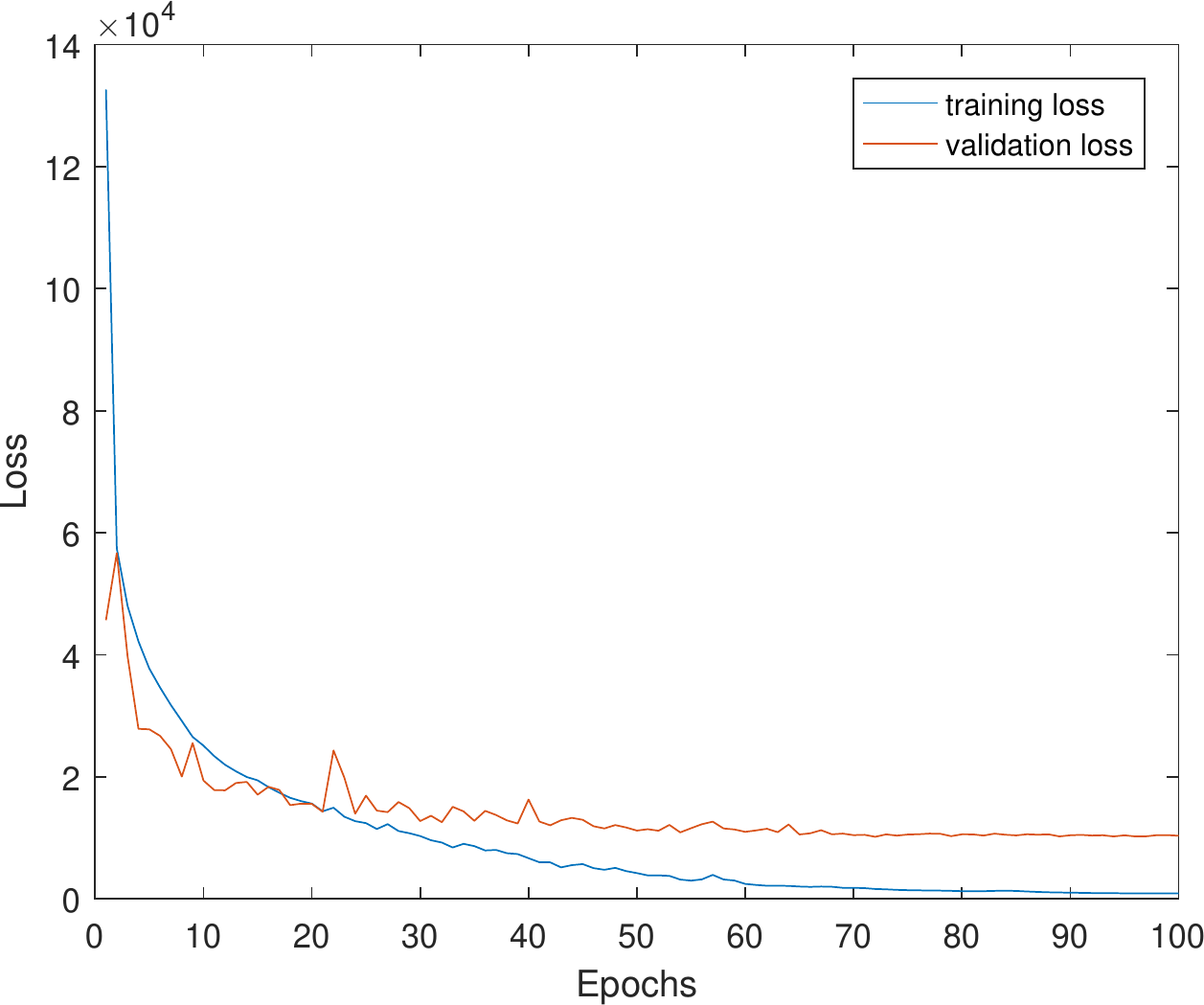}
	\end{center}
	\caption{Example of training phantoms used to train the U-net neural network and a plot of the training and validation loss vs. epoch .}
	\label{fig:training_images}
\end{figure}


\newcommand{\imwidth}{1.3in}
\begin{figure*}[!htbp]
	\begin{center}
		\begin{tabular}{@{}c@{}c@{}c@{}c@{}c@{}c@{}}
			& (1) & (2) &(3)&(4)&(5)
			\tabularnewline
			\rotatebox[origin=c]{90}{Ground Truth} & \
			\includegraphics[align = c,width=\imwidth]{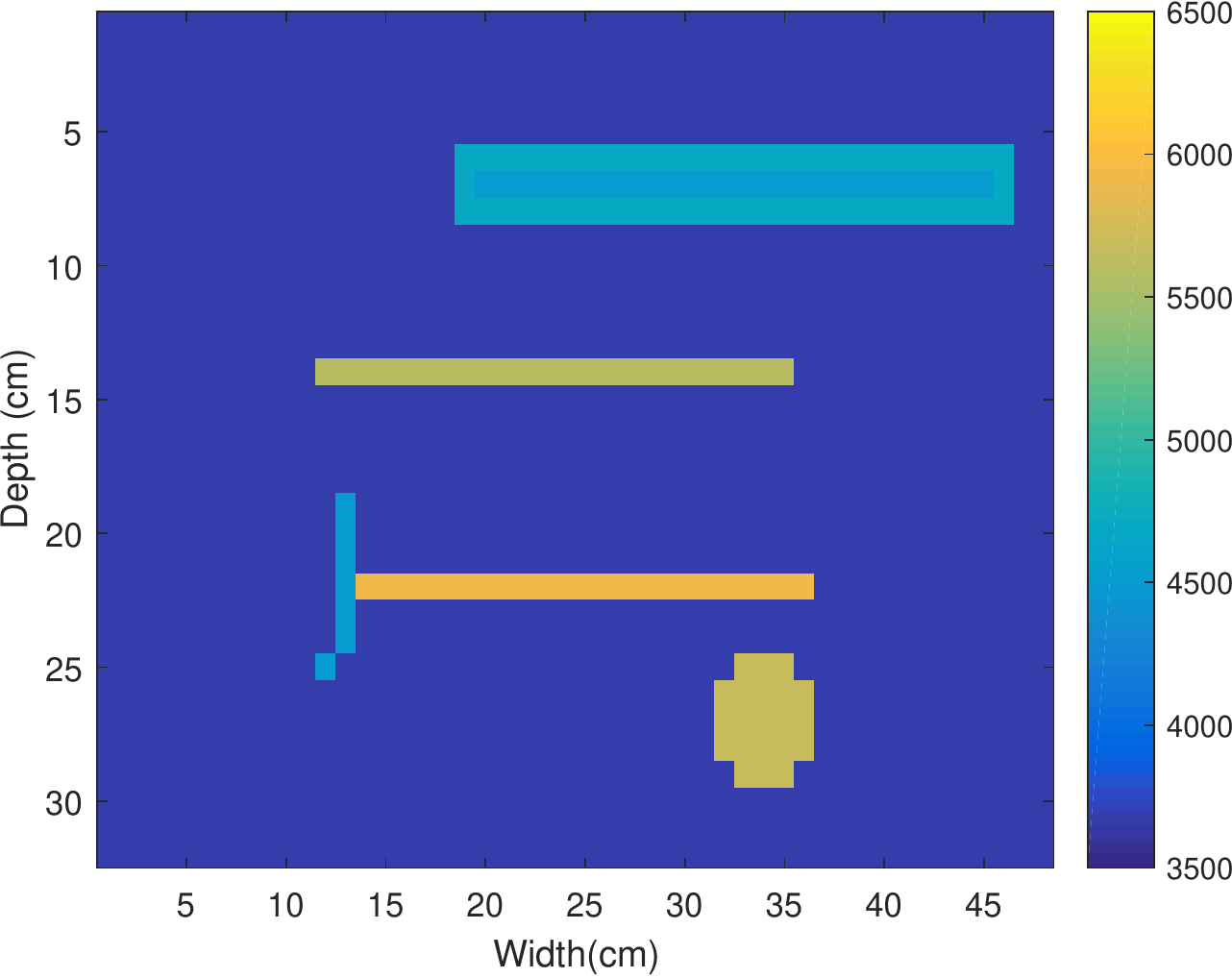}&\
			\includegraphics[align = c,width=\imwidth]{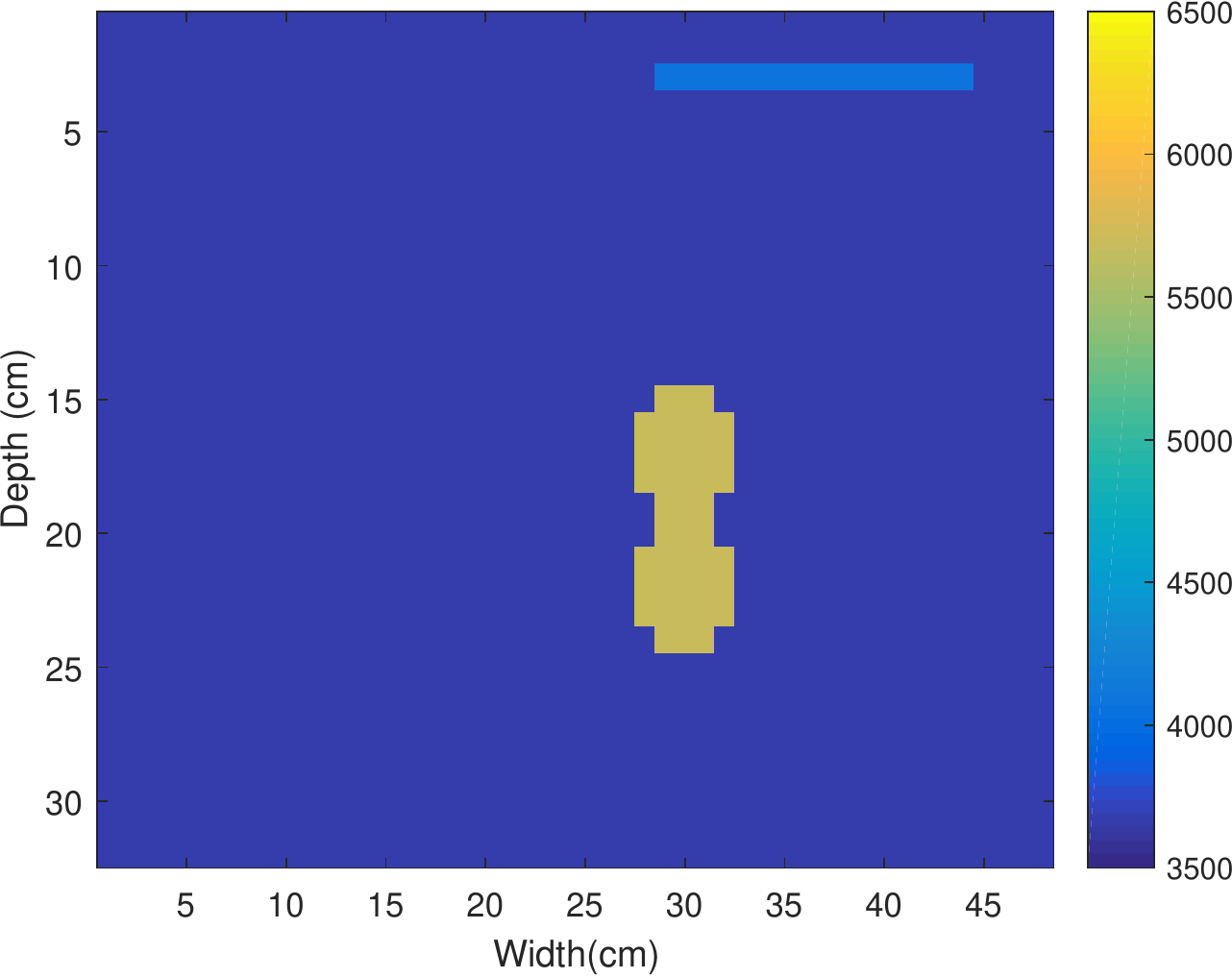}&\
			\includegraphics[align = c,width=\imwidth]{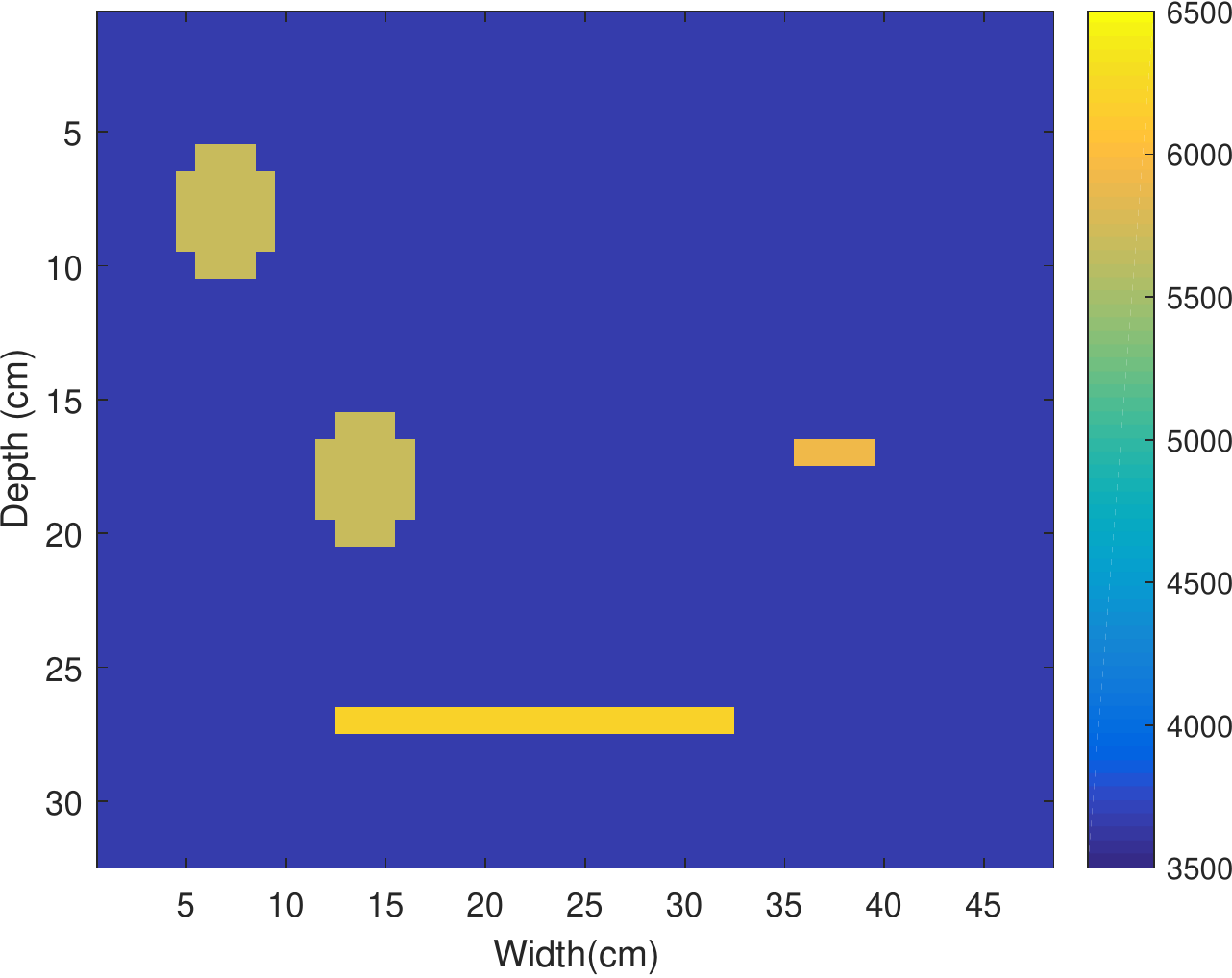}&\
			\includegraphics[align = c,width=\imwidth]{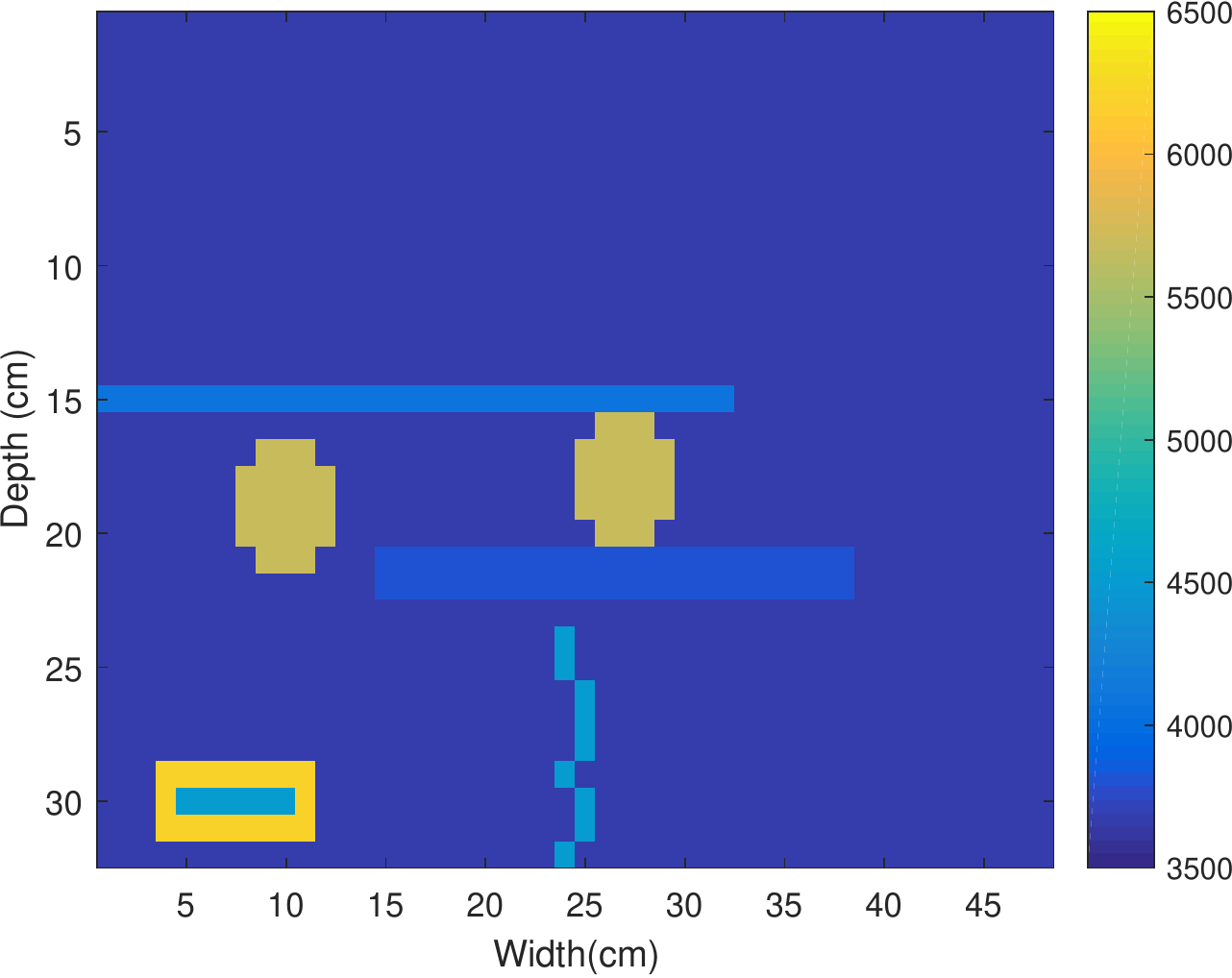}&\
			\includegraphics[align = c,width=\imwidth]{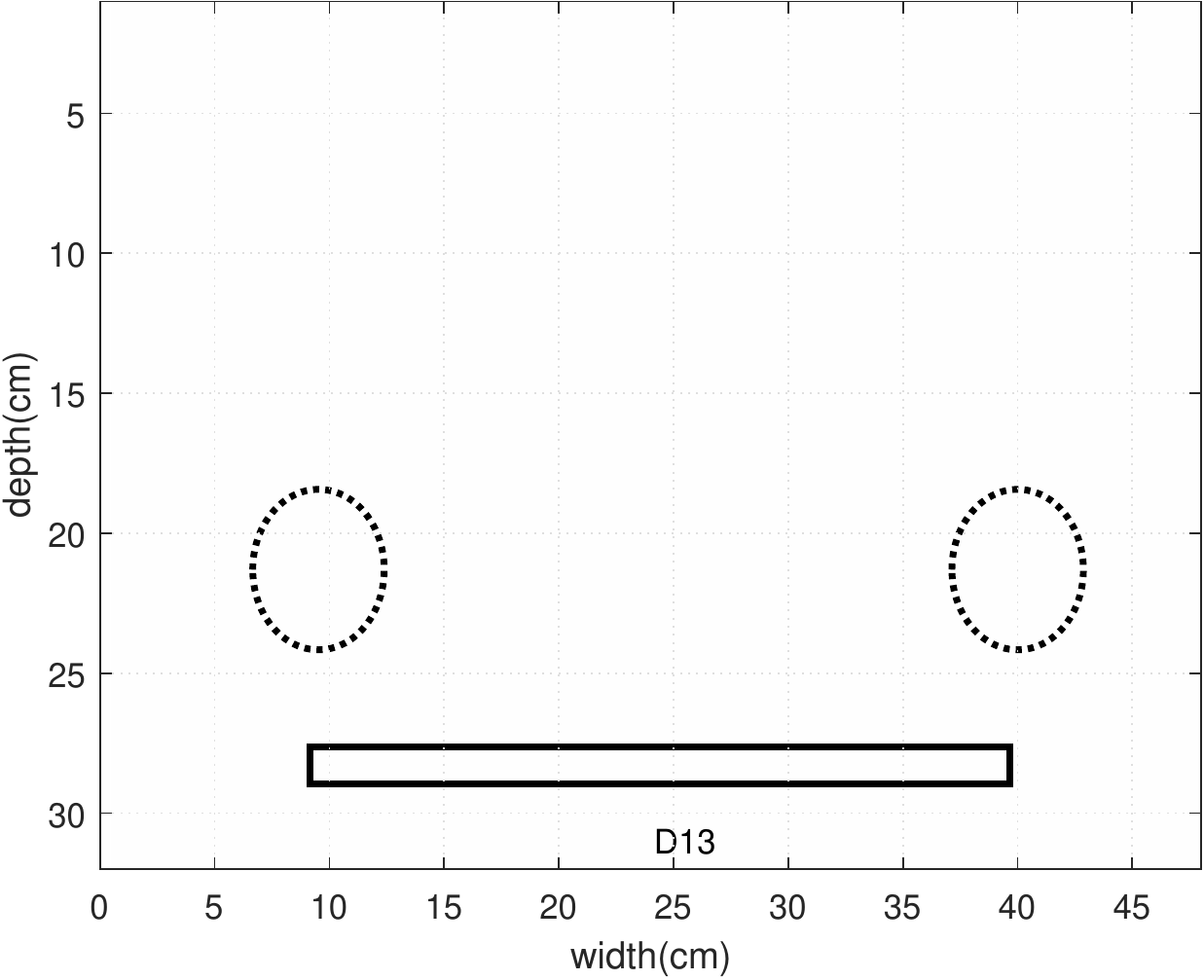}
			
			\tabularnewline
			
			\rotatebox[origin=c]{90}{SAFT} & \
			\includegraphics[align = c,width=\imwidth]{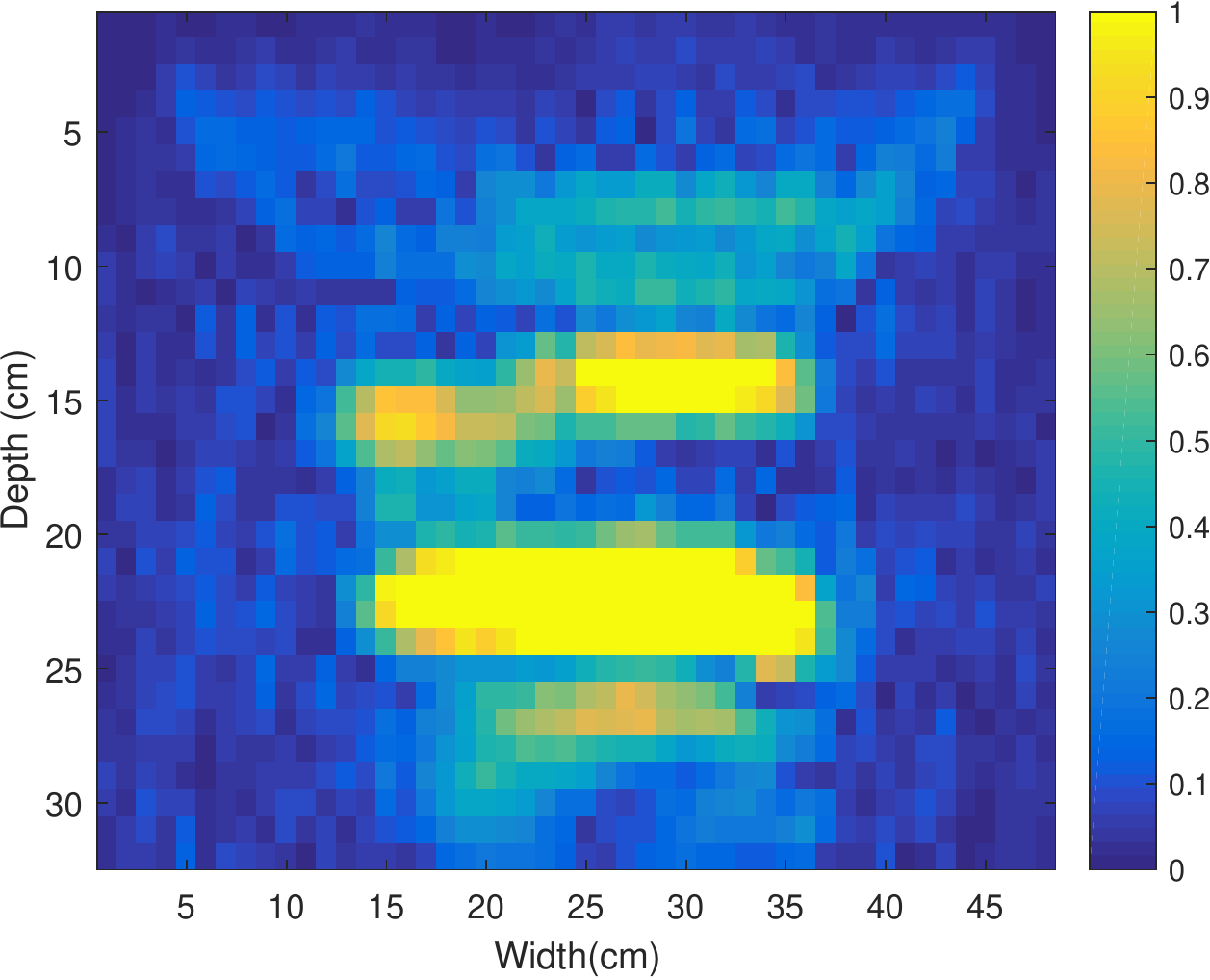}&\
			\includegraphics[align = c,width=\imwidth]{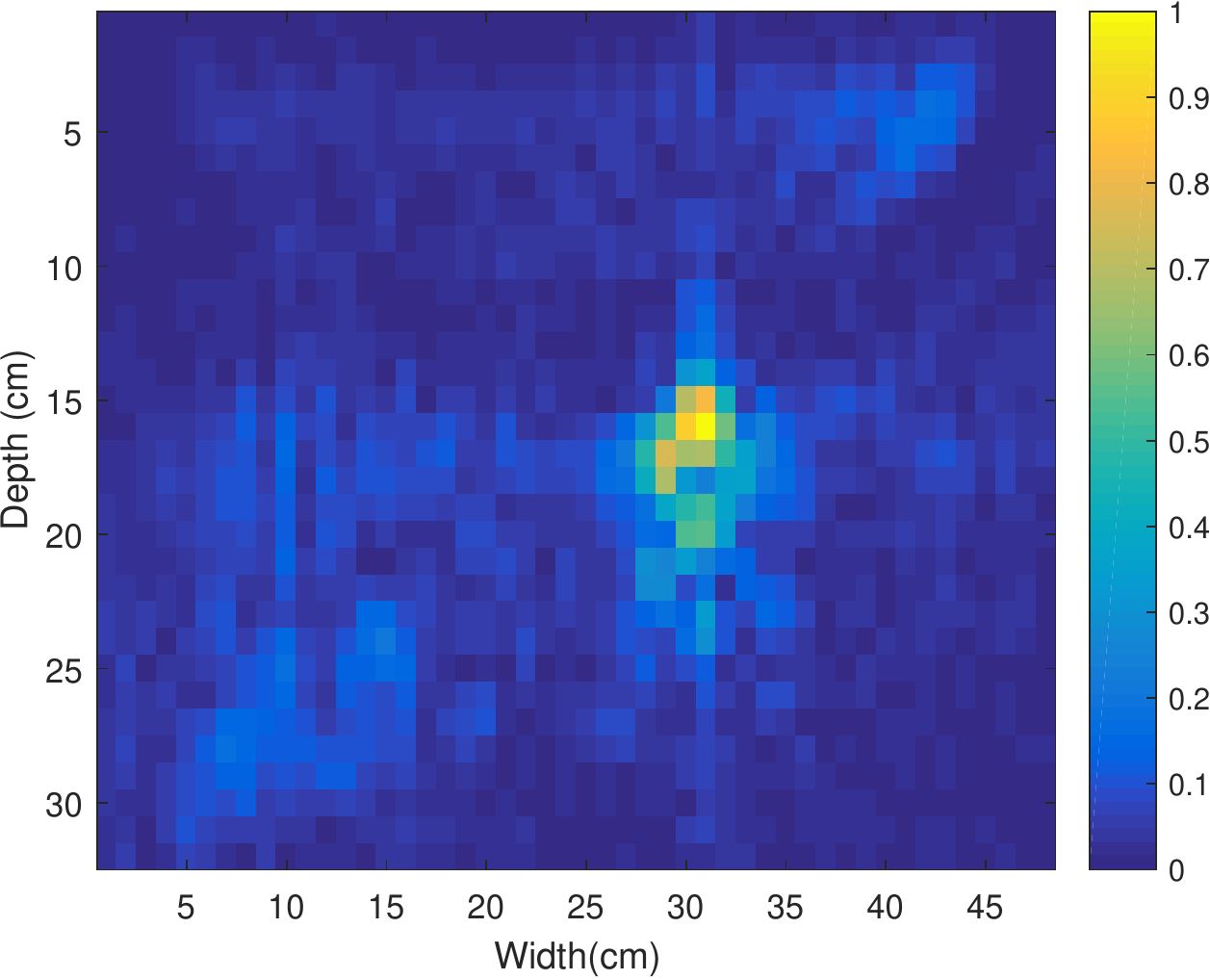}&\
			\includegraphics[align = c,width=\imwidth]{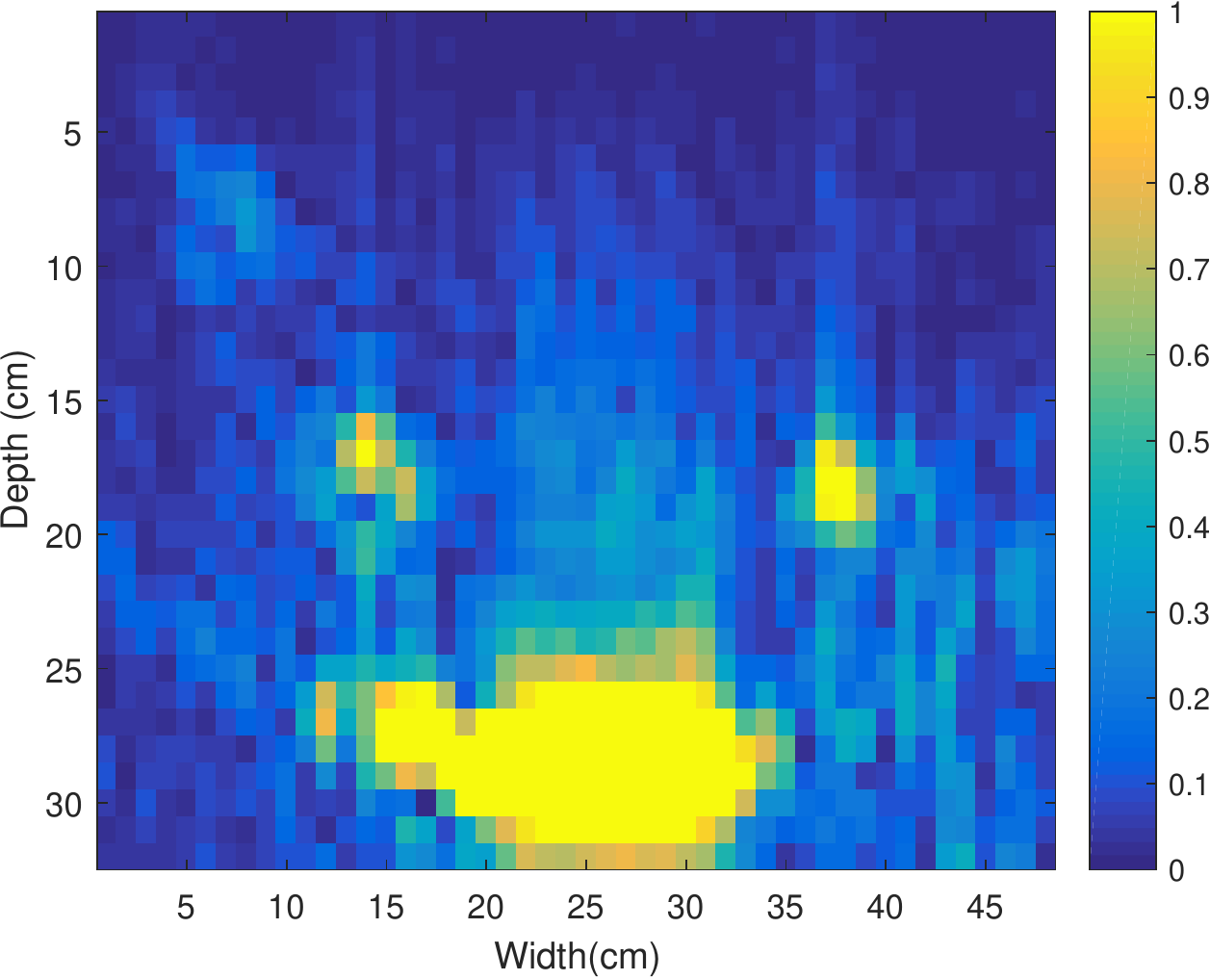}&\
			\includegraphics[align = c,width=\imwidth]{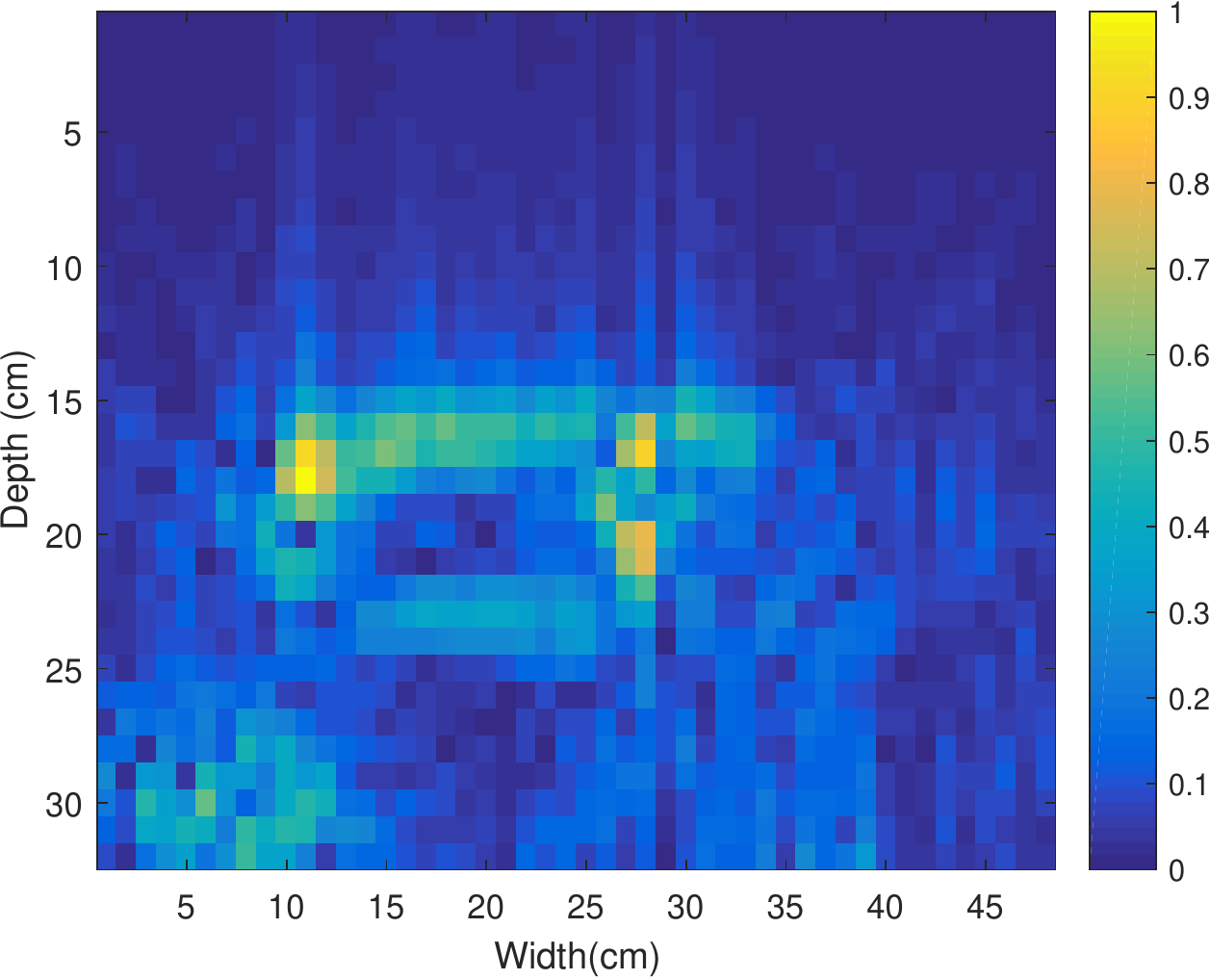}&\
			\includegraphics[align = c,width=\imwidth]{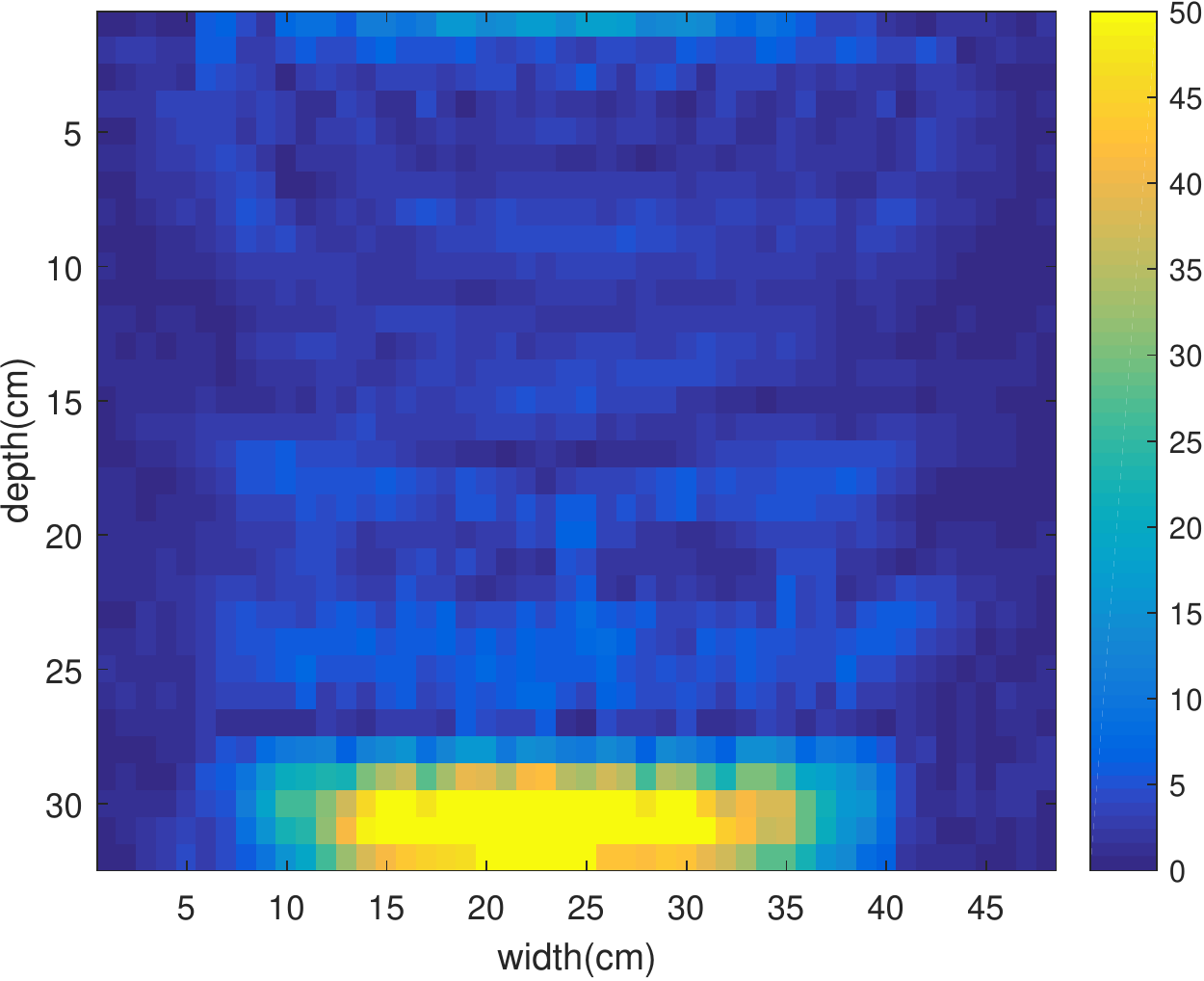}
			
			\tabularnewline
			
			\rotatebox[origin=c]{90}{L-MBIR} & \
			\includegraphics[align = c,width=\imwidth]{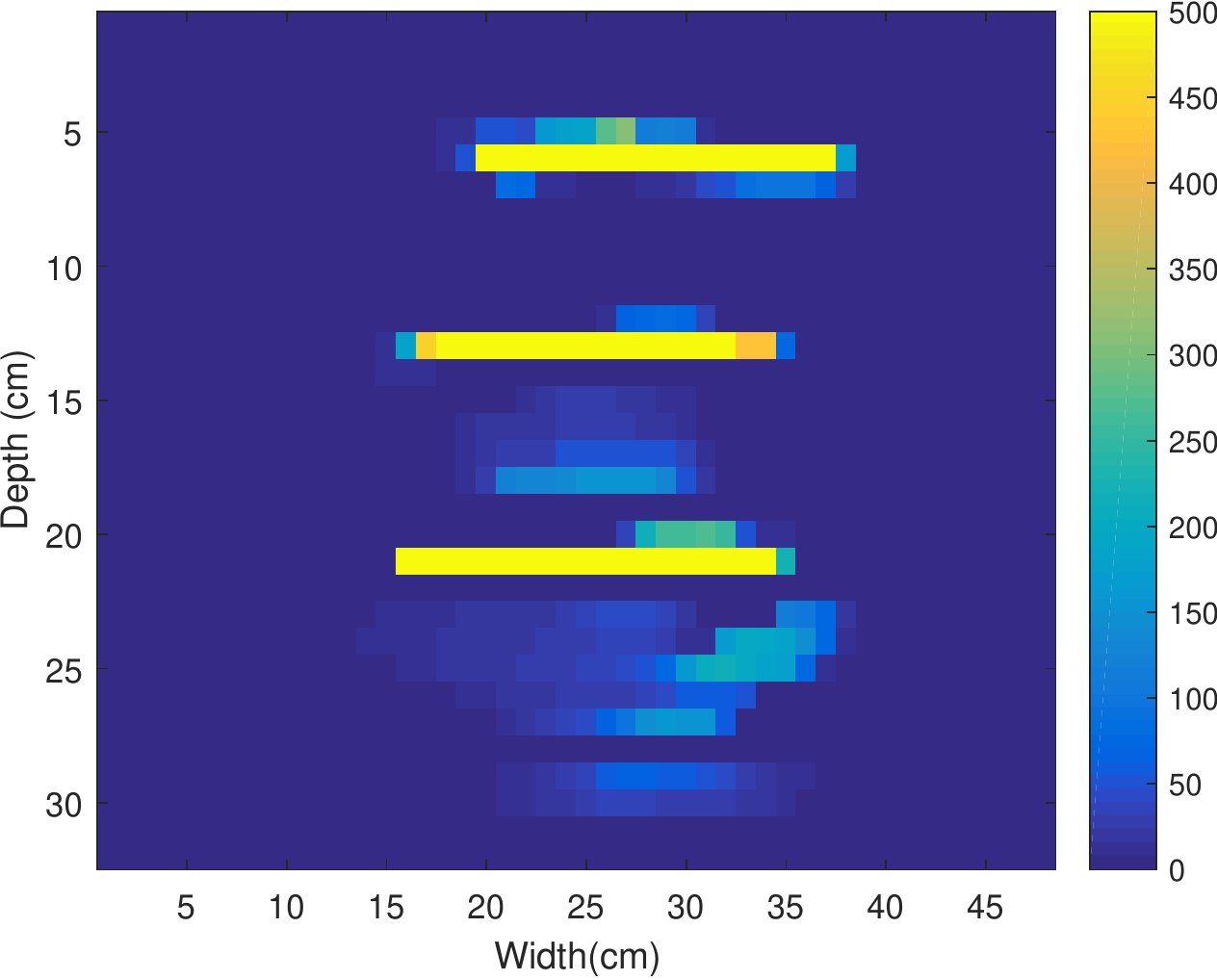}&\
			\includegraphics[align = c,width=\imwidth]{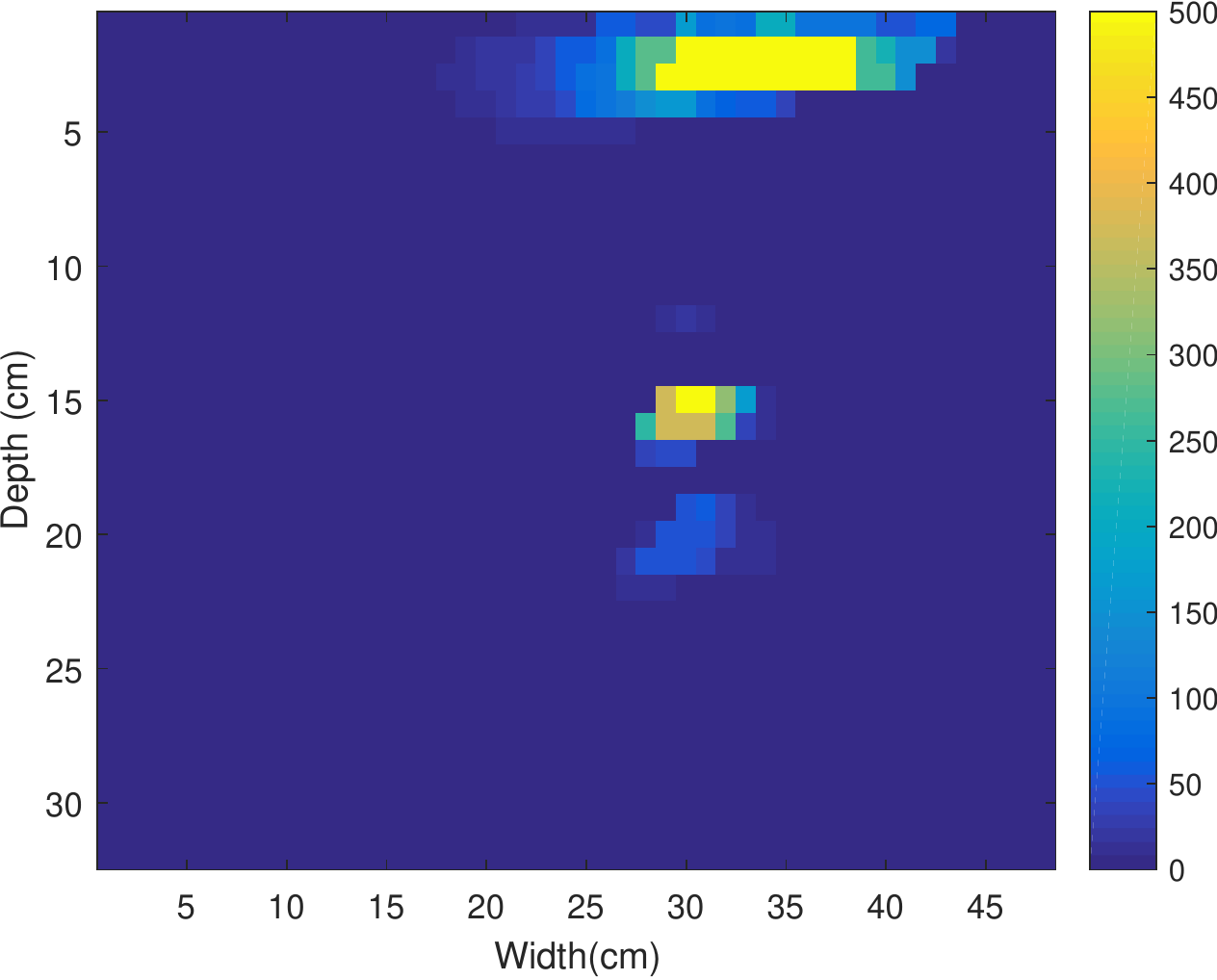}&\
			\includegraphics[align = c,width=\imwidth]{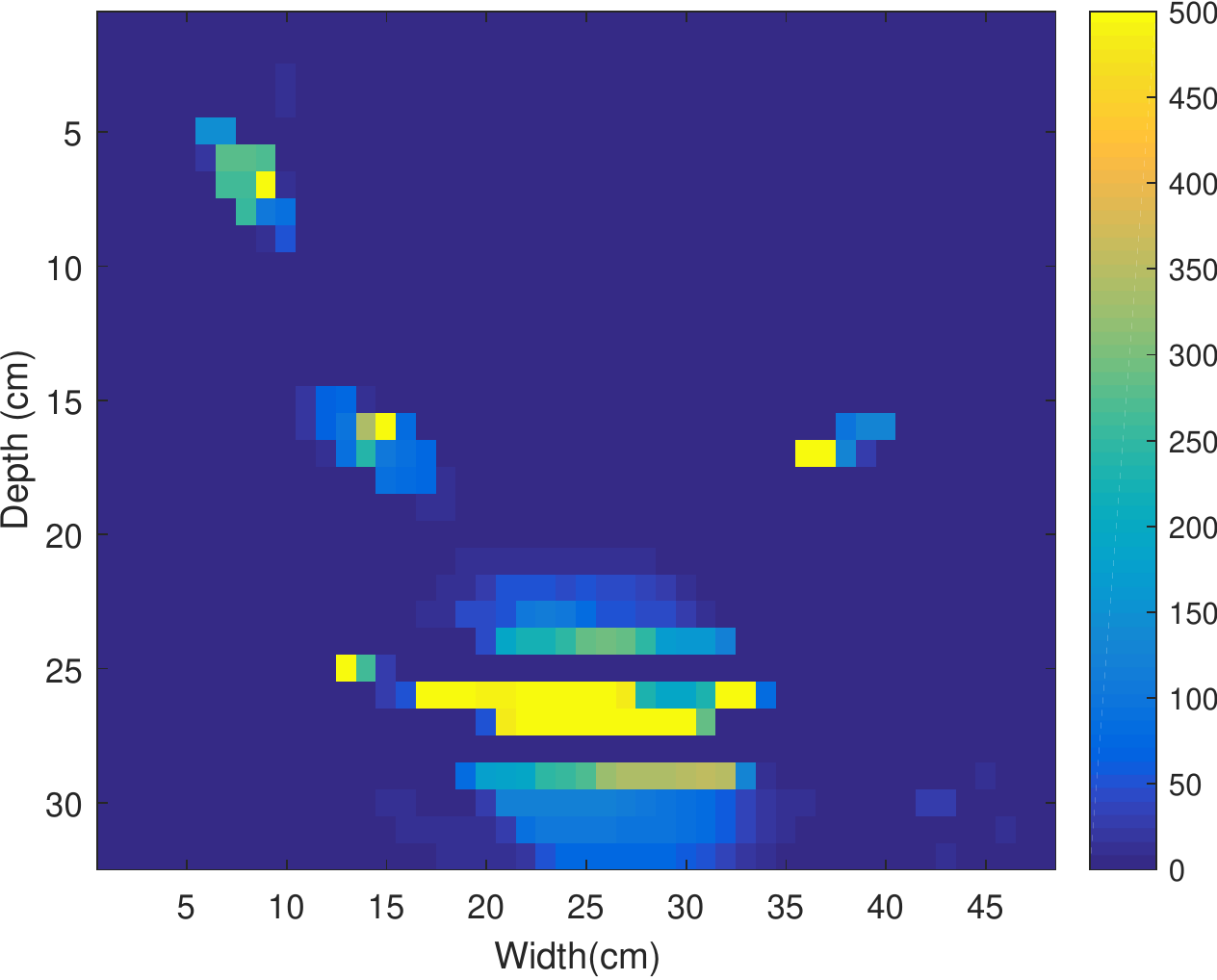}&\
			\includegraphics[align = c,width=\imwidth]{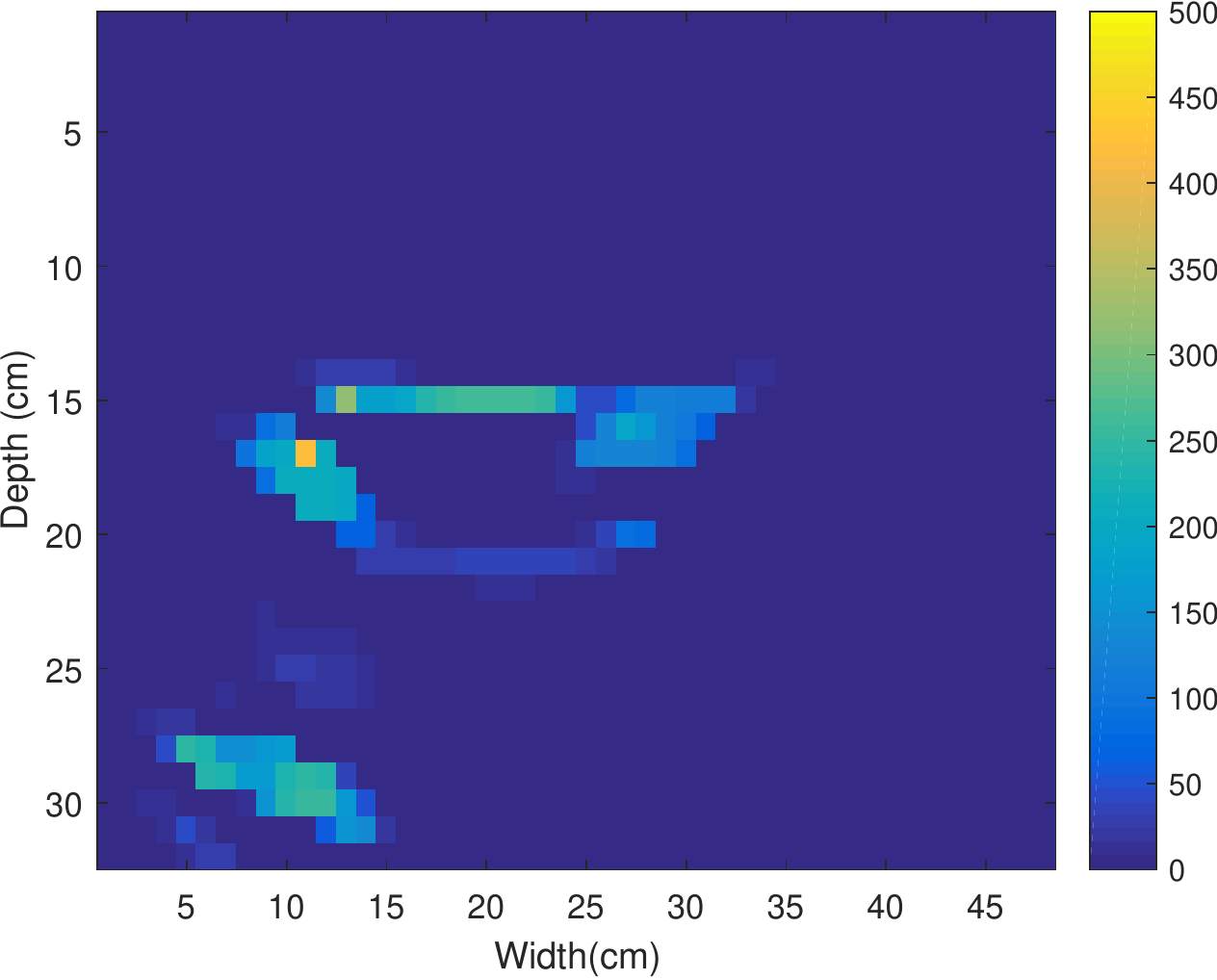}&\
			\includegraphics[align = c,width=\imwidth]{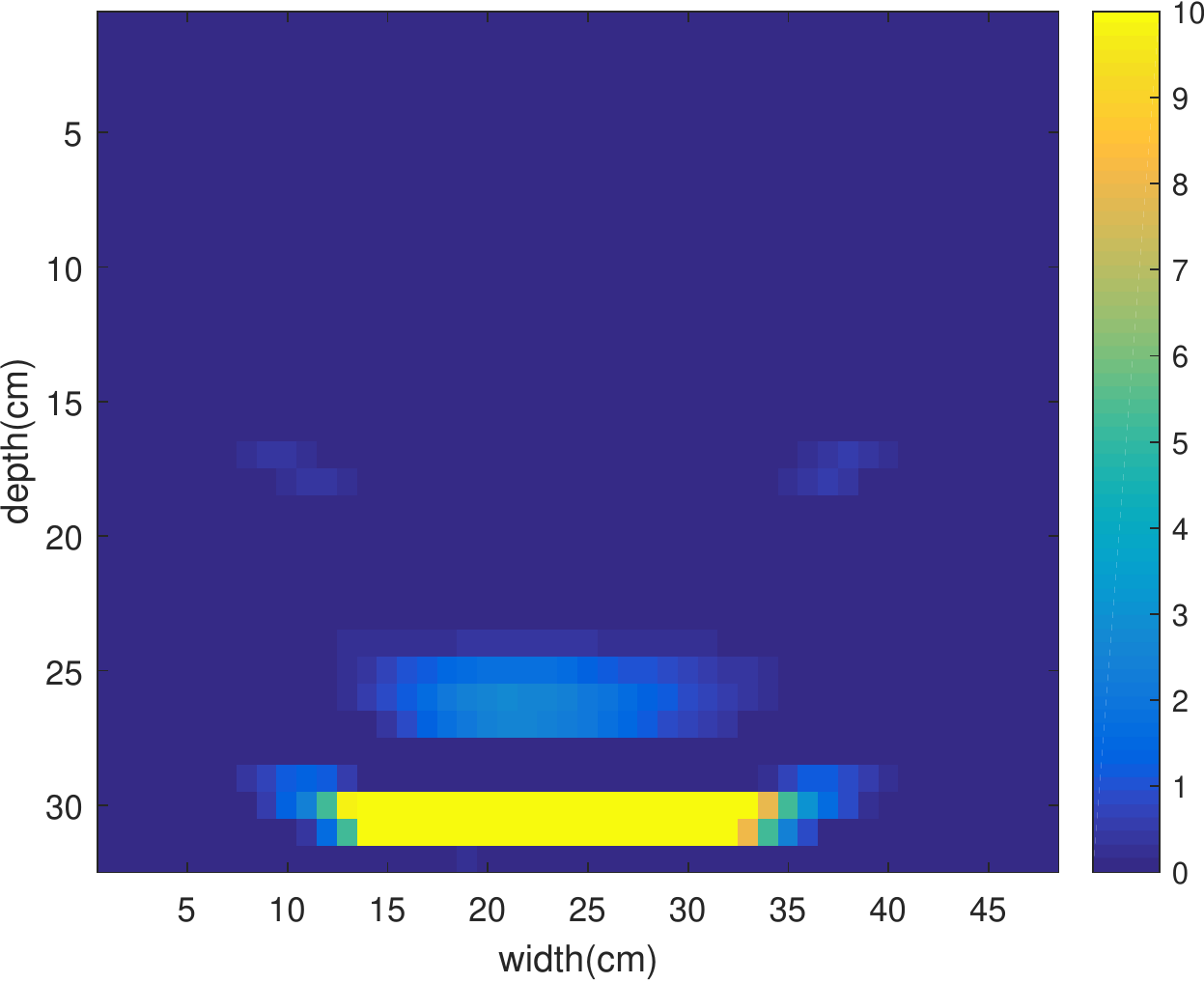}
			
			\tabularnewline
			\rotatebox[origin=c]{90}{DDL} & \
			\includegraphics[align = c,width=\imwidth]{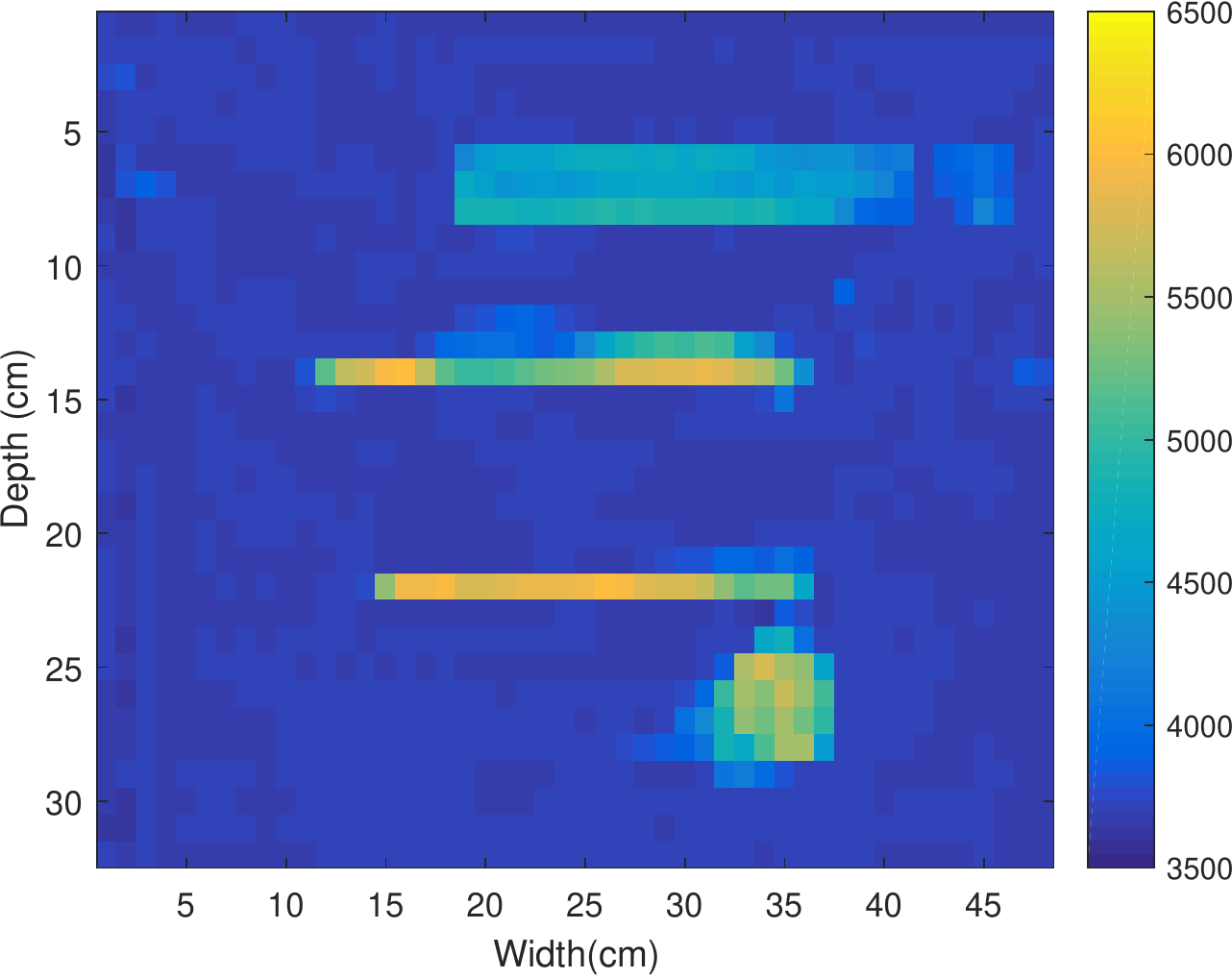}&\
			\includegraphics[align = c,width=\imwidth]{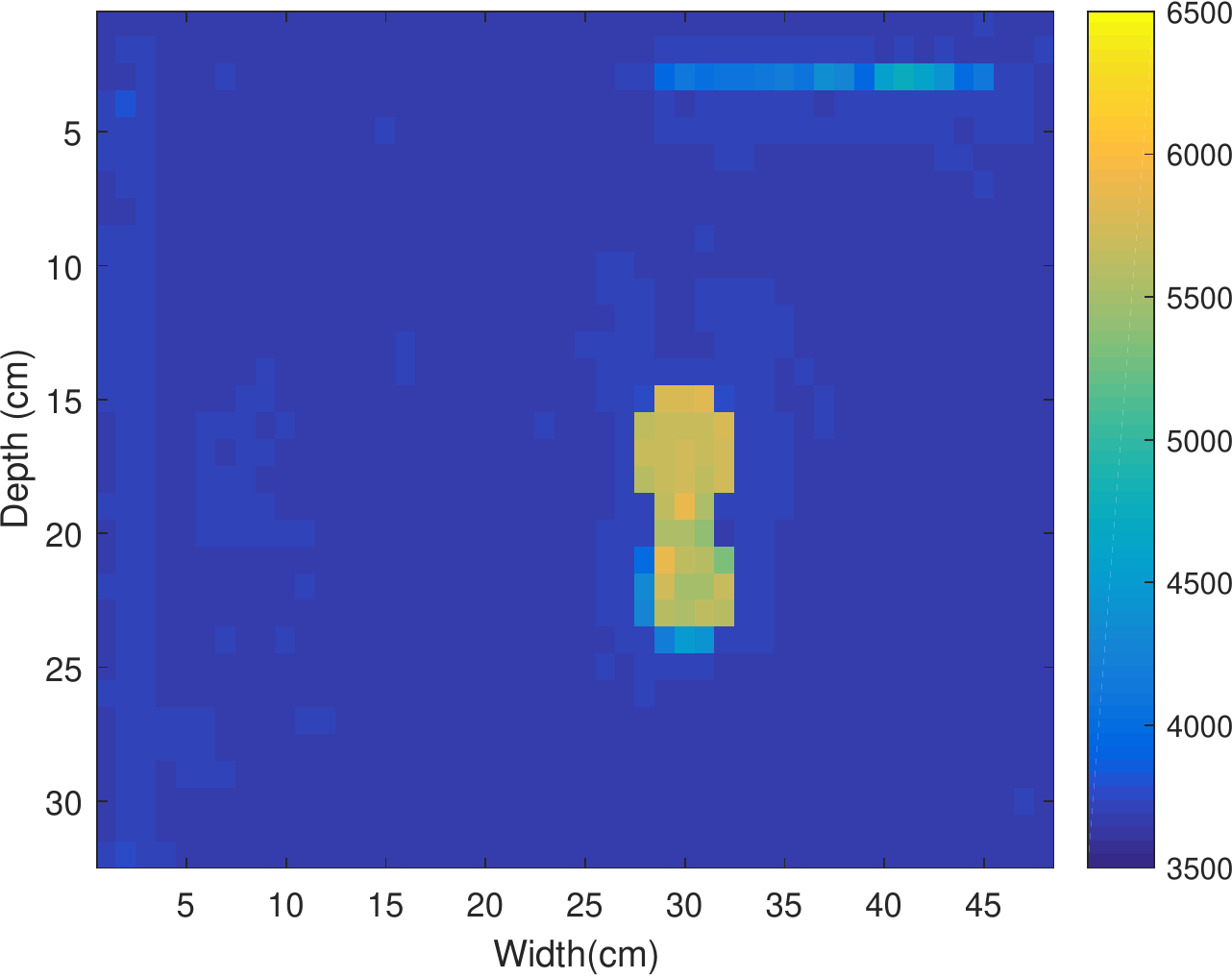}&\
			\includegraphics[align = c,width=\imwidth]{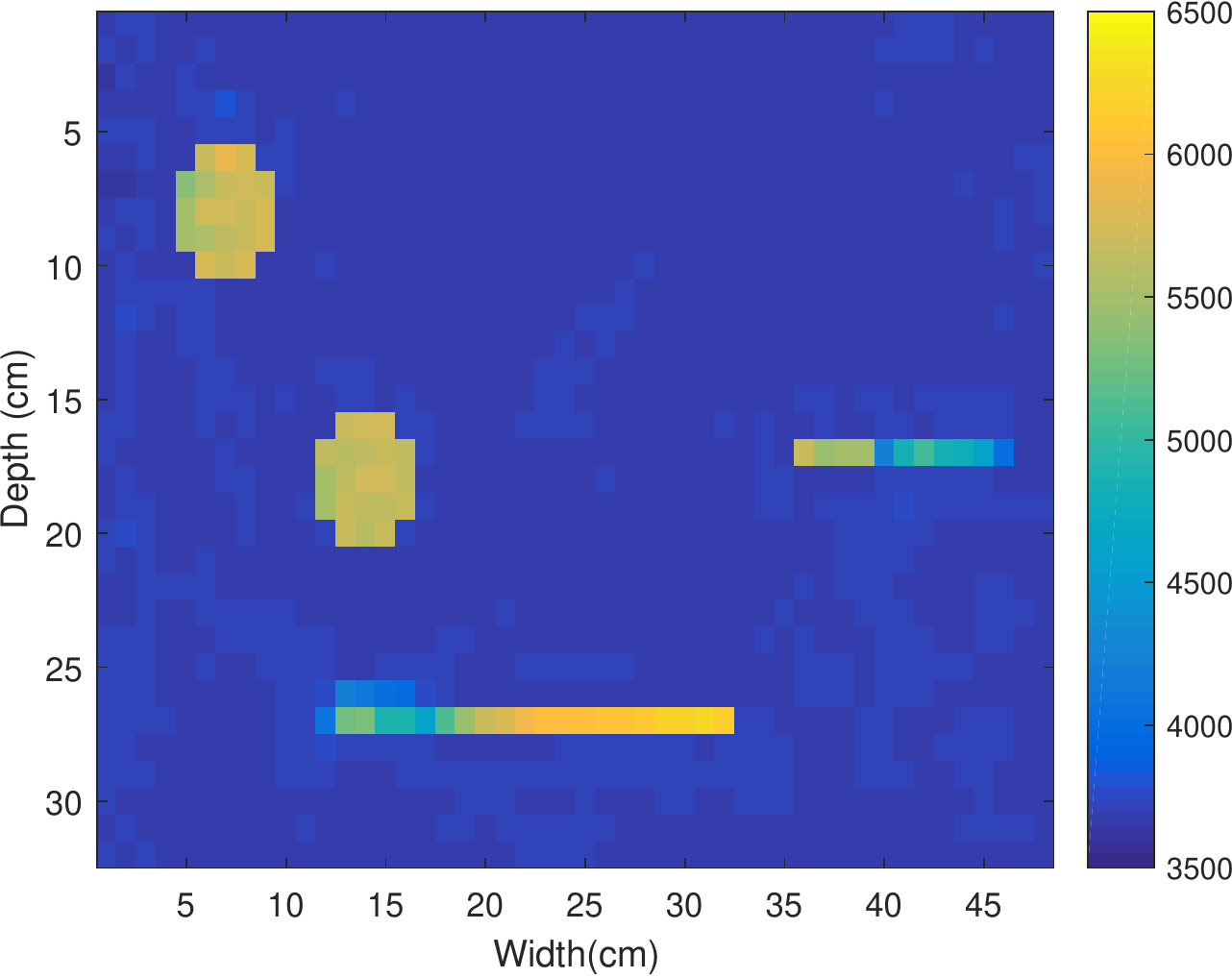}&\
			\includegraphics[align = c,width=\imwidth]{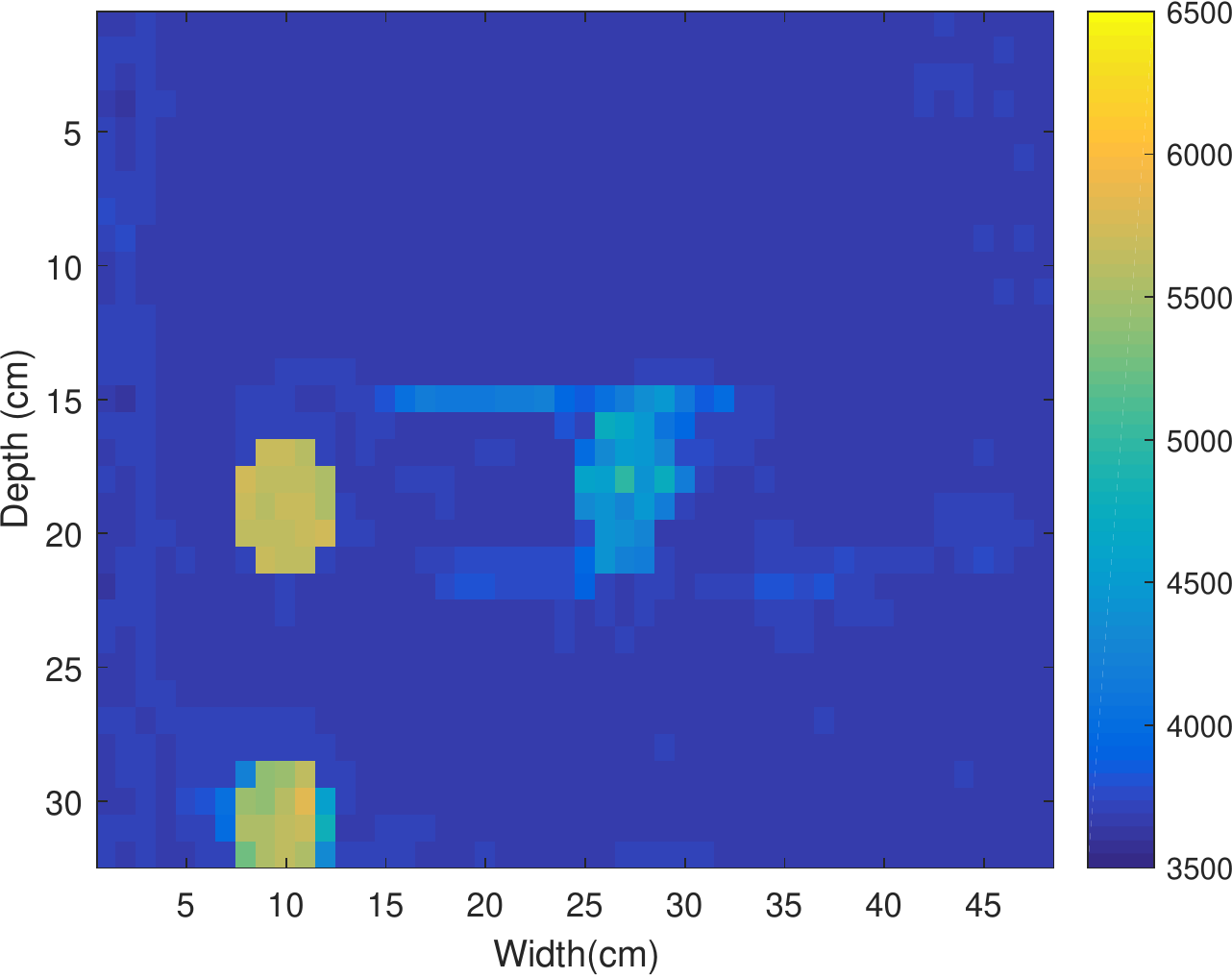}&\
			\includegraphics[align = c,width=\imwidth]{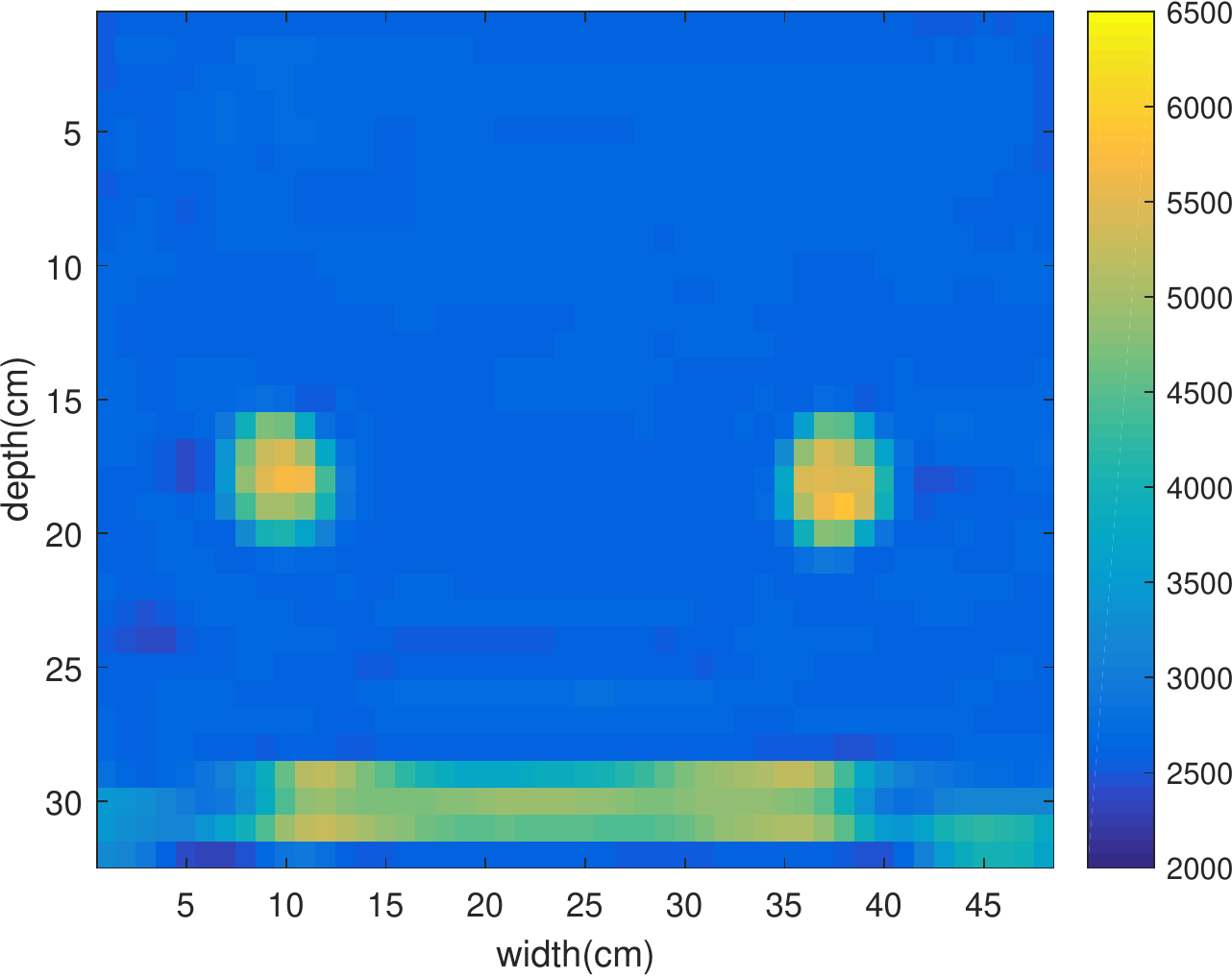}
			
			\tabularnewline
			
		\end{tabular}
	\end{center}
	\vspace{-4mm}
	\caption{
		Comparison between all reconstruction results from k-wave simulated data (samples 1 to 4) from the test set only and from experimental data (sample 5): the first row is the ground truth, the second row is SAFT reconstruction, the third row is linear MBIR reconstruction, and the fourth row is the proposed DDL reconstruction. DDL results in reconstructions with dramatic reduction in artifacts and is able to image behind occluding objects.}
	\label{fig:recon}
\end{figure*}
\setlength{\textfloatsep}{0.1cm}
\begin{table}
	\caption{Average NRMSE and SSIM for SAFT, L-MBIR, and DDL for reconstructions from the test set after a best least-squares linear fit to the ground truth.}
	\label{tab:msekwave}
	\begin{center}
		\begin{tabular}{|c|c|c|c|}
			\hline
			Method&SAFT&L-MBIR&DDL \\ \hline
			NRMSE & 0.0614 & 0.0666 &  \textbf{0.0188} \\ \hline
			SSIM & 0.5583 & 0.4147 &  \textbf{0.9340} \\ \hline
		\end{tabular}
	\end{center}
\end{table}
\setlength{\textfloatsep}{0.1cm}
We compare the proposed DDL algorithm to SAFT \cite{shao2011design} and L-MBIR \cite{almansouri2018anisotropic}. 
We used a ten transducer system with an acquisition geometry in which one of the transducers transmits while the others receive. 
The transducers are spaced 4 cm apart.
The transmitter sends a pulse of duration 50 $\mu$s with a carrier frequency of 52 KHz.
The receiver collects 263 samples with a sampling frequency of 200 KHz.
The received signals were post-processed to eliminate the direct arrival signal.

The training data set was generated by using the k-Wave simulation software with its default boundary conditions \cite{kwave2010} and is representative of the type of defects seen while inspecting thick, reinforced concrete walls  with embedded steel plates.
The density and attenuation were fixed, while the speed of sound varied from pixel to pixel depending on the material of the object. 
The background of the field of view is concrete with acoustic speed of 3680 m/s.
The steel rebar is represented as circles with speed 5660 m/s.
The defects are represented as rectangles with different speeds with possible alkali--silica reactions (ASR) \cite{barnes2014structure,hobbs1988alkali} inside with speed 4500 m/s.
The cracks are represented as ASR crooked lines.
In order to train the deep neural network, we used  $1800$ images of size $32 \times 48$ pixels for training the network, $200$ for validation and $200$ for testing. 
Stochastic gradient descent is used to optimize the loss function with batch size = 1, learning rate = 0.0001, and momentum = 0.5.
The optimization was performed using the PyTorch \cite{pyTorch} library.
Fig.~\ref{fig:training_images} shows examples of the training phantoms used to generate the ultrasound training data along with the curves for training and validation plots for the data-set. 

Samples 1 to 4 in Fig.~\ref{fig:recon} shows reconstructed images from the test set (not used in training) using SAFT, the linear MBIR of \eqref{eq:us_lin} and the proposed DDL approach. 
Notice that the units of each method are different, i.e. the unit in SAFT, L-MBIR, and DDL are pressure, reflectivity, and speed of sound, respectively.
What makes DDL advantageous is that we are reconstructing the same unit as the ground truth which makes it easy to interpret the image.
Also, notice that while L-MBIR is qualitatively superior to the SAFT reconstruction, it is unable to resolve some of the artifacts caused by reverberations and shadowing due to the linear model for the reflected signal shown in Fig.~\ref{fig:recon} in sample 1 and 2, respectively.
In contrast, the proposed DDL approach suppresses these artifacts and results in dramatic improvements in image quality. 
However, for some weak reflections, DDL-generated artifacts are similar to the features in the training set (e.g. circular objects). 
		Such artifacts may be hard to spot and pass as actual features in the specimen, such as the bottom left object in sample 4. 

Sample 5 in Fig.~\ref{fig:recon} shows reconstructed images from experimental data. 
The experiment is described in detail in \cite{almansouri2018anisotropic}. 
The training of DDL was done with the same k-wave simulated data,
except the concrete acoustic speed was changed to (2620 m/s), 
i.i.d. Gaussian noise, $\mathcal{N}(0,200^2)$, was added to the ground truth to account for the modeling error, 
and the direct arrival signals were not eliminated.
Notice that the DDL reconstruction significantly improves reconstruction quality by accurately reconstructing the steel rebar as well as the plate compared to SAFT and L-MBIR.
	

Table \ref{tab:msekwave} shows the NRMSE and SSIM of the three approaches and illustrates that the proposed method also results in significant improvement of the quantitative accuracy of the results.
A least squares fit to ground truth is used to scale and shift each reconstruction to optimize the RMSE for each method as in \cite{gupta2018cnn}.  NRMSE uses $\|x_r-x_g\|/\|x_g\|,$ where $x_g$ is ground truth and $x_r$ is the best fit of the reconstruction to $x_g$. For SSIM, the ground truth and the best fit reconstruction are both converted to image intensity using Matlab mat2gray with the same intensity scale for each.

\vspace{-2mm}
\section{Conclusions \label{sec:Concl}}

In this paper, we proposed a method for reflection model ultrasound reconstruction using a deep neural network. 
Our algorithm obtains an initial estimate using a linear back projection and then uses a trained neural network to map this preliminary reconstruction to the final solution. 
Using simulated and experimental data we showed that our algorithm produces a dramatic improvement in reconstruction quality compared to the typically used analytic algorithms as well as iterative algorithms based on linear models.

\vspace{-2mm}
\section{Acknowledgment}
Hani Almansouri and C.A. Bouman were supported by the U.S. Department of Energy. 
G.T. Buzzard was partially supported by NSF CCF-1763896. 
S.Venkatakrishnan and Hector Santos-Villalobos were supported by the U.S. Department of Energy’s staff office of the Under Secretary for Science and Energy under the Subsurface Technology and Engineering Research, Development, and Demonstration (SubTER) Crosscut program, and the office of Nuclear Energy under the Light Water Reactor Sustainability (LWRS) program. 


\bibliographystyle{IEEEtran}
{
\footnotesize
\bibliography{ultrasound}

\begin{thebibliography}{10}
\providecommand{\url}[1]{#1}
\csname url@samestyle\endcsname
\providecommand{\newblock}{\relax}
\providecommand{\bibinfo}[2]{#2}
\providecommand{\BIBentrySTDinterwordspacing}{\spaceskip=0pt\relax}
\providecommand{\BIBentryALTinterwordstretchfactor}{4}
\providecommand{\BIBentryALTinterwordspacing}{\spaceskip=\fontdimen2\font plus
\BIBentryALTinterwordstretchfactor\fontdimen3\font minus
  \fontdimen4\font\relax}
\providecommand{\BIBforeignlanguage}[2]{{%
\expandafter\ifx\csname l@#1\endcsname\relax
\typeout{** WARNING: IEEEtran.bst: No hyphenation pattern has been}%
\typeout{** loaded for the language `#1'. Using the pattern for}%
\typeout{** the default language instead.}%
\else
\language=\csname l@#1\endcsname
\fi
#2}}
\providecommand{\BIBdecl}{\relax}
\BIBdecl

\bibitem{hoegh2015extended}
K.~Hoegh and L.~Khazanovich, ``Extended synthetic aperture focusing technique
  for ultrasonic imaging of concrete,'' \emph{NDT \& E International}, vol.~74,
  pp. 33--42, 2015.

\bibitem{bernard2017ultrasonic}
S.~Bernard, V.~Monteiller, D.~Komatitsch, and P.~Lasaygues, ``Ultrasonic
  computed tomography based on full-waveform inversion for bone quantitative
  imaging,'' \emph{Physics in Medicine \& Biology}, vol.~62, no.~17, p. 7011,
  2017.

\bibitem{shao2011}
Z.~Shao, L.~Shi, Z.~Shao, and J.~Cai, ``{Design and application of a small size
  {SAFT} imaging system for concrete structure},'' \emph{Review of Scientific
  Instruments}, vol.~82, no.~7, p. 073708, 2011.

\bibitem{Engle2014}
B.~J. Engle, J.~L.~W. Schmerr, and A.~Sedov, ``{Quantitative ultrasonic phased
  array imaging},'' \emph{AIP Conf. Proc.}, vol. 1581, no.~7, p.~49, 2014.

\bibitem{dobie2013}
G.~Dobie, S.~G. Pierce, and G.~Hayward, ``{The feasibility of synthetic
  aperture guided wave imaging to a mobile sensor platform},'' \emph{NDT and E
  International}, vol.~58, no.~7, pp. 10--17, 2013.

\bibitem{almansouri2018anisotropic}
H.~Almansouri, S.~Venkatakrishnan, D.~Clayton, Y.~Polsky, C.~Bouman, and
  H.~Santos-Villalobos, ``Anisotropic modeling and joint-map stitching for
  improved ultrasound model-based iterative reconstruction of large and thick
  specimens,'' in \emph{AIP Conference Proceedings}, vol. 1949, no.~1.\hskip
  1em plus 0.5em minus 0.4em\relax AIP Publishing, 2018, p. 030002.

\bibitem{mccann2017convolutional}
M.~T. McCann, K.~H. Jin, and M.~Unser, ``Convolutional neural networks for
  inverse problems in imaging: A review,'' \emph{IEEE Signal Processing
  Magazine}, vol.~34, no.~6, pp. 85--95, 2017.

\bibitem{han2017deep}
Y.~Han and J.~C. Ye, ``Deep residual learning approach for sparse-view {CT}
  reconstruction,'' in \emph{Fully Three-Dimensional Image Reconstruction in
  Radiology and Nuclear Medicine}.\hskip 1em plus 0.5em minus 0.4em\relax
  Fully3D conference organization, 2017.

\bibitem{jin2017deep}
K.~H. Jin, M.~T. McCann, E.~Froustey, and M.~Unser, ``Deep convolutional neural
  network for inverse problems in imaging,'' \emph{IEEE Transactions on Image
  Processing}, vol.~26, no.~9, pp. 4509--4522, 2017.

\bibitem{han2018deep}
Y.~Han, J.~Yoo, H.~H. Kim, H.~J. Shin, K.~Sung, and J.~C. Ye, ``Deep learning
  with domain adaptation for accelerated projection-reconstruction {MR},''
  \emph{Magnetic resonance in medicine}, vol.~80, no.~3, pp. 1189--1205, 2018.

\bibitem{wang2016accelerating}
S.~Wang, Z.~Su, L.~Ying, X.~Peng, S.~Zhu, F.~Liang, D.~Feng, and D.~Liang,
  ``Accelerating magnetic resonance imaging via deep learning,'' in
  \emph{Biomedical Imaging (ISBI), 2016 IEEE 13th International Symposium
  on}.\hskip 1em plus 0.5em minus 0.4em\relax IEEE, 2016, pp. 514--517.

\bibitem{antholzer2017deep}
S.~Antholzer, M.~Haltmeier, and J.~Schwab, ``Deep learning for photoacoustic
  tomography from sparse data,'' \emph{arXiv preprint arXiv:1704.04587}, 2017.

\bibitem{hauptmann2018model}
A.~Hauptmann, F.~Lucka, M.~Betcke, N.~Huynh, J.~Adler, B.~Cox, P.~Beard,
  S.~Ourselin, and S.~Arridge, ``Model based learning for accelerated,
  limited-view {3D} photoacoustic tomography,'' \emph{IEEE transactions on
  medical imaging}, 2018.

\bibitem{mousavi2017learning}
A.~Mousavi and R.~G. Baraniuk, ``Learning to invert: Signal recovery via deep
  convolutional networks,'' in \emph{Acoustics, Speech and Signal Processing
  (ICASSP), 2017 IEEE International Conference on}.\hskip 1em plus 0.5em minus
  0.4em\relax IEEE, 2017, pp. 2272--2276.

\bibitem{sun2018efficient}
Y.~Sun, Z.~Xia, and U.~S. Kamilov, ``Efficient and accurate inversion of
  multiple scattering with deep learning,'' \emph{Optics express}, vol.~26,
  no.~11, pp. 14\,678--14\,688, 2018.

\bibitem{venkatakrishnan2013plug}
S.~V. Venkatakrishnan, C.~A. Bouman, and B.~Wohlberg, ``Plug-and-play priors
  for model based reconstruction,'' in \emph{Global Conference on Signal and
  Information Processing (GlobalSIP), 2013 IEEE}.\hskip 1em plus 0.5em minus
  0.4em\relax IEEE, 2013, pp. 945--948.

\bibitem{sreehari2016plug}
S.~Sreehari, S.~V. Venkatakrishnan, B.~Wohlberg, G.~T. Buzzard, L.~F. Drummy,
  J.~P. Simmons, and C.~A. Bouman, ``Plug-and-play priors for bright field
  electron tomography and sparse interpolation,'' \emph{IEEE Transactions on
  Computational Imaging}, vol.~2, no.~4, pp. 408--423, 2016.

\bibitem{zhang2017learning}
K.~Zhang, W.~Zuo, S.~Gu, and L.~Zhang, ``Learning deep cnn denoiser prior for
  image restoration,'' in \emph{2017 IEEE Conference on Computer Vision and
  Pattern Recognition (CVPR)}, July 2017, pp. 2808--2817.

\bibitem{gupta2018cnn}
H.~Gupta, K.~H. Jin, H.~Q. Nguyen, M.~T. McCann, and M.~Unser, ``{CNN}-based
  projected gradient descent for consistent {CT} image reconstruction,''
  \emph{IEEE transactions on medical imaging}, vol.~37, no.~6, pp. 1440--1453,
  2018.

\bibitem{rick2017one}
J.~Rick~Chang, C.-L. Li, B.~Poczos, B.~Vijaya~Kumar, and A.~C.
  Sankaranarayanan, ``One network to solve them all--solving linear inverse
  problems using deep projection models,'' in \emph{Proceedings of the IEEE
  Conference on Computer Vision and Pattern Recognition}, 2017, pp. 5888--5897.

\bibitem{adler2018learned}
J.~Adler and O.~{\"O}ktem, ``Learned primal-dual reconstruction,'' \emph{IEEE
  transactions on medical imaging}, vol.~37, no.~6, pp. 1322--1332, 2018.

\bibitem{meinhardt17learning}
\BIBentryALTinterwordspacing
T.~Meinhardt, M.~Möller, C.~Hazirbas, and D.~Cremers, ``Learning proximal
  operators: Using denoising networks for regularizing inverse imaging
  problems,'' in \emph{ICCV}, October 2017. [Online]. Available:
  \url{https://github.com/tum-vision/learn_prox_ops}
\BIBentrySTDinterwordspacing

\bibitem{kwave2010}
B.~E. Treeby and B.~T. Cox, ``k-{W}ave: Matlab toolbox for the simulation and
  reconstruction of photoacoustic wave fields,'' \emph{Journal of biomedical
  optics}, vol.~15, no.~2, p. 021314, 2010.

\bibitem{thibault2007three}
J.-B. Thibault, K.~D. Sauer, C.~A. Bouman, and J.~Hsieh, ``A three-dimensional
  statistical approach to improved image quality for multislice helical ct,''
  \emph{Medical physics}, vol.~34, no.~11, pp. 4526--4544, 2007.

\bibitem{ronneberger2015u}
O.~Ronneberger, P.~Fischer, and T.~Brox, ``U-net: {C}onvolutional networks for
  biomedical image segmentation,'' in \emph{International Conference on Medical
  image computing and computer-assisted intervention}.\hskip 1em plus 0.5em
  minus 0.4em\relax Springer, 2015, pp. 234--241.

\bibitem{shao2011design}
Z.~Shao, L.~Shi, Z.~Shao, and J.~Cai, ``Design and application of a small size
  saft imaging system for concrete structure,'' \emph{Review of Scientific
  Instruments}, vol.~82, no.~7, p. 073708, 2011.

\bibitem{barnes2014structure}
P.~Barnes and J.~Bensted, \emph{Structure and performance of cements}.\hskip
  1em plus 0.5em minus 0.4em\relax CRC Press, 2014.

\bibitem{hobbs1988alkali}
D.~W. Hobbs, \emph{Alkali-silica reaction in concrete}.\hskip 1em plus 0.5em
  minus 0.4em\relax London, 1988.

\bibitem{pyTorch}
A.~Paszke, S.~Gross, S.~Chintala, G.~Chanan, E.~Yang, Z.~DeVito, Z.~Lin,
  A.~Desmaison, L.~Antiga, and A.~Lerer, ``Automatic differentiation in
  pytorch,'' 2017.

\end{thebibliography}
}

\end{document}